\documentclass[aps,prl,amsmath,amssymb,twocolumn,floatfix,superscriptaddress,showpacs]{revtex4-1}
\usepackage{setspace}
\usepackage{graphicx}
\usepackage{dcolumn}
\usepackage{epsfig}
\usepackage{color}

\begin{document}

\title{Observation of {D}irac surface states in the noncentrosymmetric superconductor {BiPd}}

\author{H. M. Benia}
\email{h.benia@fkf.mpg.de}
\affiliation{Max-Planck-Institut f\"ur Festk\"orperforschung, Heisenbergstrasse 1, D-70569 Stuttgart, Germany}
\author{E. Rampi}
\affiliation{Department of Physics \& Astronomy, University of British Columbia, Vancouver, British Columbia V6T~1Z1, Canada}
\author{C. Trainer}
\affiliation{SUPA, School of Physics and Astronomy, University of St Andrews, North Haugh, St Andrews, Fife KY16 9SS, UK}
\author{C. M. Yim}
\affiliation{SUPA, School of Physics and Astronomy, University of St Andrews, North Haugh, St Andrews, Fife KY16 9SS, UK}
\author{A. Maldonado}
\affiliation{SUPA, School of Physics and Astronomy, University of St Andrews, North Haugh, St Andrews, Fife KY16 9SS, UK}
\author{D. C. Peets}
\affiliation{Max-Planck-Institut f\"ur Festk\"orperforschung, Heisenbergstrasse 1, D-70569 Stuttgart, Germany}
\affiliation{Advanced Materials Laboratory, Fudan University, Shanghai 200438, China}
\author{A. St\"ohr}
\affiliation{Max-Planck-Institut f\"ur Festk\"orperforschung, Heisenbergstrasse 1, D-70569 Stuttgart, Germany}
\author{U. Starke}
\affiliation{Max-Planck-Institut f\"ur Festk\"orperforschung, Heisenbergstrasse 1, D-70569 Stuttgart, Germany}
\author{K. Kern}
\affiliation{Max-Planck-Institut f\"ur Festk\"orperforschung, Heisenbergstrasse 1, D-70569 Stuttgart, Germany}
\affiliation{Institut de Physique, Ecole Polytechnique F\'{e}d\'{e}rale de Lausanne, 1015 Lausanne, Switzerland}
\author{A. Yaresko}
\affiliation{Max-Planck-Institut f\"ur Festk\"orperforschung, Heisenbergstrasse 1, D-70569 Stuttgart, Germany}
\author{G. Levy}
\affiliation{Department of Physics \& Astronomy, University of British Columbia, Vancouver, British Columbia V6T~1Z1, Canada}
\affiliation{Quantum Matter Institute, University of British Columbia, Vancouver, British Columbia V6T~1Z4, Canada}
\author{A. Damascelli}
\affiliation{Department of Physics \& Astronomy, University of British Columbia, Vancouver, British Columbia V6T~1Z1, Canada}
\affiliation{Quantum Matter Institute, University of British Columbia, Vancouver, British Columbia V6T~1Z4, Canada}
\author{C. R. Ast}
\affiliation{Max-Planck-Institut f\"ur Festk\"orperforschung, Heisenbergstrasse 1, D-70569 Stuttgart, Germany}
\author{A. P. Schnyder}
\affiliation{Max-Planck-Institut f\"ur Festk\"orperforschung, Heisenbergstrasse 1, D-70569 Stuttgart, Germany}
\author{P. Wahl}
\email{wahl@st-andrews.ac.uk}
\affiliation{SUPA, School of Physics and Astronomy, University of St Andrews, North Haugh, St Andrews, Fife KY16 9SS, UK}
\affiliation{Max-Planck-Institut f\"ur Festk\"orperforschung, Heisenbergstrasse 1, D-70569 Stuttgart, Germany}

\date{\today}

\begin{abstract}
Materials with strong spin-orbit coupling (SOC) have in recent years become a subject of intense research due to their potential applications in spintronics and quantum information technology. In particular, in systems which break inversion symmetry, SOC facilitates the Rashba-Dresselhaus effect, leading to a lifting of spin degeneracy in the bulk and intricate spin textures of the Bloch wave functions.
Here, by combining angular resolved photoemission (ARPES) and low temperature scanning tunneling microscopy (STM) measurements with relativistic first-principles band structure calculations, we examine the role of SOC in single crystals of noncentrosymmetric BiPd. We report the detection of several Dirac surface states, one of which exhibits an extremely large spin splitting. Unlike the surface states in inversion-symmetric systems, the Dirac surface states of BiPd have completely different properties at opposite faces of the crystal and are not trivially linked by symmetry. The spin-splitting of the surface states exhibits a strong anisotropy by itself, which can be linked to the low in-plane symmetry of the surface termination.
\end{abstract}

\maketitle


The interplay of strong spin-orbit coupling (SOC) with superconductivity has become a major focus of research in
recent years, as both are essential ingredients to stabilize Majorana bound states. The spin-orbit interaction affects the electronic states in a material in various ways and in particular can lead to non-trivial topologies of the band structure. In topological insulators SOC separates the conduction and valence bands, leading to an insulating state with an inverted band gap \cite{hasanKaneRMP,qiZhangRMP,hasanMooreAnnRev}. The latter leads directly to the presence of 
Dirac surface states protected by time-reversal symmetry \cite{D.Hsieh13022009,XiaNatPhys09,seoYazdaniNat10}.
Another consequence of SOC is the Rashba effect \cite{rashba1960,astPRL07,crepaldiGrioniPRL12,bahramyNatComm12}, which in the absence of inversion symmetry lifts the spin degeneracy of the electronic bands, generating intricate spin textures in the electronic wave functions \cite{hsiehHasanNature09,IshizakaNatMat11,wangGedikPRL11}. Commonly observed at surfaces or interfaces, in noncentrosymmetric materials the Rashba-Dresselhaus effect leads to a lifting of spin-degeneracy of the {\slshape bulk} bands. Combined with superconductivity this can lead to mixing of spin-singlet and spin-triplet pairing components \cite{bauerSigristBook,bauer2004heavy} and, more interestingly, to a topologically nontrivial superconducting phase \cite{Schnyder2008gfback,beriPRB10,Andreas_flat,satoFujimotoPRB09}.

Noncentrosymmetric BiPd \cite{kheiker1953x,Zhuravlev1957,bhatt1980kristallstruktur,Bhatt1979P17,Ionov1989} becomes superconducting below $3.8$~K \cite{Alekseevskii1952,joshi2011superconductivity,mondal2012andreev,matano2013nmr,sunNatComm15,Peets2016} and
offers a unique opportunity to study the interplay between SOC and superconductivity. The large spin-orbit interaction of the heavy element Bi results in a sizeable spin splitting of the bulk bands of BiPd~\cite{sunNatComm15}. This in turn can lead to nontrivial wavefunction topologies and unconventional superconducting states~\cite{sasakiAndoPRL11,levyStroscioPRL13}. 
Along with the half-Heusler compounds~\cite{Liu2011,Kim2016,Nakajima2016} and PbTaSe$_2$~\cite{Ali2014,Bian2016,Guan2016}, BiPd constitutes a rare example of a noncentrosymmetric superconductor which cleaves easily, enabling high-resolution surface-sensitive spectroscopy of its electronic states\cite{sunNatComm15,Neupane1505}.

In this paper we report the observation of Rashba spin-split Dirac surface states of noncentrosymmetric BiPd by
angle-resolved photoemission spectroscopy (ARPES) and low-temperature scanning tunneling microscopy and spectroscopy (STM/STS). Due to the lack of inversion symmetry, the (010) and (0\=10) surface states can appear at different energies and exhibit different dispersions and spin-polarizations.
By combining the experimental results with relativistic first-principles band structure calculations we identify the Dirac surface states of both the (010) and (0\=10) surfaces. This observation of distinct Dirac surface states originating from the opposing surface terminations represents a unique demonstration of the impact of the lack of inversion symmetry on the electronic states.

\begin{figure*}[t!]
\begin{center}
\includegraphics[width=0.8\textwidth]{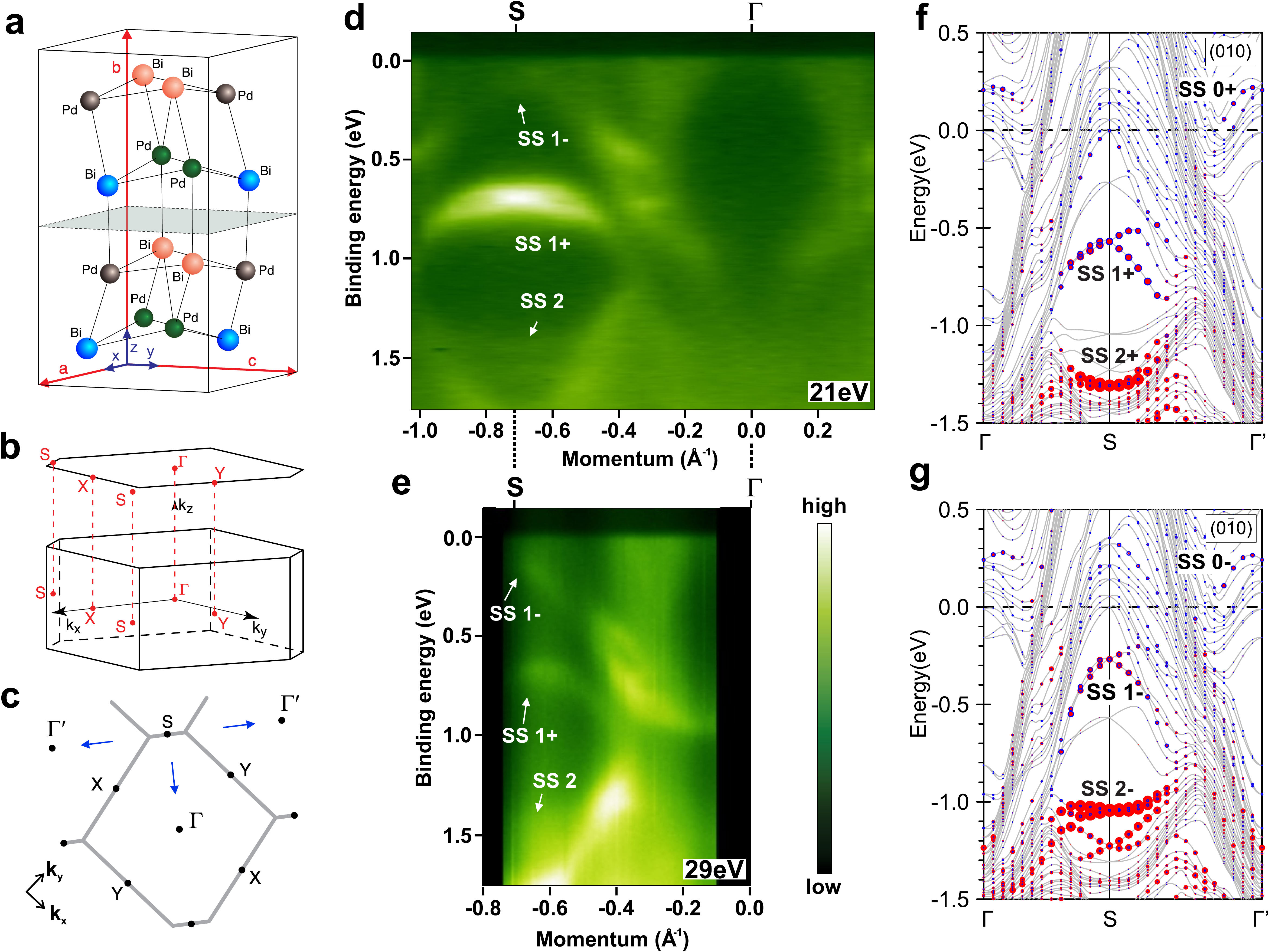}
\end{center}
\caption{\label{mFig1}(a) Crystal structure of BiPd, showing the preferred cleaving plane. (b) Schematic representation of the Brillouin zone of BiPd as well as the surface Brillouin zone of the (010) and (0\=10) surfaces. (c) Surface Brillouin zone with the cuts shown in panels (d)-(g) in blue.
(d) Experimental electronic band structure of a BiPd(010) surface along the S--$\Gamma$ direction ($\nu=21~\mathrm{eV}$). (e) Electronic structure measured with $\nu=29~\mathrm{eV}$. The photoemission intensities in (d) and (e) and in other photoemission maps in this paper are displayed using the colour scale shown in e.
(f) and (g) Calculated electronic structure of BiPd in slab geometry including the cuts shown in (d) and (e). The size of the circles is proportional to the spectral weight of Bi $6p$ states in the first (red) and second (blue) layer of the (010) and (0$\bar{\textrm{1}}$0) surfaces.
The surface states are labelled SS$n$+ and SS$n$-, where + and - denote whether they occur on the (010) or (0$\bar{\textrm{1}}$0) surface, while $n$ numbers the surface states sequentially with increasing binding energy.}
\end{figure*}

The crystal growth using a modified Bridgman--Stockbarger technique has been described in detail elsewhere~\cite{PeetsLT27}. The crystals were cooled slowly through the $\alpha-\beta$ phase transition to maximize the domain size of the low-temperature $\alpha$ phase; resulting in high-quality crystals~\cite{Peets2016}.  At low temperature $\alpha$-BiPd (in the following referred to as ``BiPd'') forms in the noncentrosymmetric space group P2$_1$ \cite{Zhuravlev1957,bhatt1980kristallstruktur,Bhatt1979P17,Ionov1989}. The structure is characterized by two double layers stacked along the monoclinic $b$ axis, which are related by a 180$^{\circ}$ screw symmetry [see Fig.~\ref{mFig1}(a)]. Since the bonding between double layers is weaker than within them, the crystals readily cleave perpendicular to the monoclinic $b$ axis and, as previously demonstrated~\cite{sunNatComm15}, are twinned such that both (010) and (0\=10) surfaces can appear on the same side of the crystal (see ref.~\onlinecite{supp} for details on the cleaving procedure).

ARPES measurements were performed on freshly cleaved surfaces using (i) a Helium source ($\nu=21.2~\mathrm{eV}$ and 40.8~eV) with a hemispherical SPECS HSA3500 electron analyzer, and (ii) linearly-polarized synchrotron light from the UE112-PGM undulator beamline at BESSY II with a Scienta R8000 analyzer. The sample was held at temperatures lower than 100~K during cleaving and throughout the measurements.

STM experiments were performed in a home-built low-temperature STM operating at temperatures down to $1.5~\mathrm{K}$ in cryogenic vacuum~\cite{Singh2013}. Samples were prepared by {\it in-situ} cleaving at low temperatures. Tips were cut from a PtIr wire. Bias voltages were applied to the sample. Differential conductance spectra have been recorded through a lock-in amplifier ($f=408~\mathrm{Hz}$, $V_\mathrm{mod}=2~\mathrm{mV}$).




Figures~\ref{mFig1}(b) and (c) show schematically the bulk Brillouin zone and its surface projection. Figures~\ref{mFig1}(d) and (e) show the results of ARPES, measured along the $\Gamma$--S direction in the Brillouin zone at two different photon energies. The most prominent feature of the surface electronic structure when measured with a He-I lamp is the appearance of a strong state (labelled SS1+ in Fig.~\ref{mFig1}(d)) at the S-point at 0.7~eV binding energy. In addition, at higher photon energy (Fig.~\ref{mFig1}(e)), within the same directional band gap at the S-point a surface state SS1-- can be identified, albeit with much weaker intensity.  These are identified as surface states through their lack of dispersion with varying the incident photon energy and hence $k_z$ \cite{supp}.
To understand the origin and topological nature of these surface states, we have employed fully relativistic linear muffin tin orbital calculations~\cite{andersenPRB75,book:AHY04,perlovYaresko} using a repeated slab system consisting of six BiPd double layers separated by two empty double layers which represent the vacuum. We find that around the Fermi energy $E_{\mathrm{F}}$, all the bands are mainly of Bi~$6p$ orbital character with subdominant but non-negligible contributions of Pd~$4d$ states.
The strong atomic SOC of Bi induces a spin splitting of the bands of the order of tens of meV and, moreover, results in a large energy shift of states that have predominant $p_{1/2}$ orbital character\cite{MPK80}. The latter leads to formation of a band gap at the $\Gamma$ point~\cite{sunNatComm15,MPK80}.
In Figs.~\ref{mFig1}(f) and (g), we show the calculated dispersions near $E_{\mathrm{F}}$ of the (010) and (0\=10) surfaces of BiPd, respectively, along high symmetry directions of the surface BZ [Fig.~\ref{mFig1}(b) and (c)]. The momentum-resolved surface densities of states at the (010) and (0\=10) sides are indicated by filled circles. Interestingly, Dirac surface states appear both at the $S$  and $\Gamma$  points of the surface BZ.
Thus by comparison with band structure calculations, the features SS1+ and SS1- seen in ARPES can be directly associated with the surface states of the BiPd surface.
The simultaneous observation of SS1+ and SS1- in the measurement is not reproduced in the calculations: the two states originate from opposite surface terminations, with the one at higher binding energy arising from the (010) termination and the one closer to the Fermi energy from the (0\=10) termination. Since these two terminations correspond to opposite surfaces of a single crystal, their simultaneous observation by ARPES indicates twin domains with opposite direction of the crystallographic $b$ axis within the beam spot.  A structural transition around 200$^\circ$C~\cite{Bhatt1979P17,Ionov1989} is known to cause twinning, and this type of twin boundary has been previously observed by STM~\cite{sunNatComm15}.

\begin{figure}[htb]
\includegraphics[width=0.9\columnwidth]{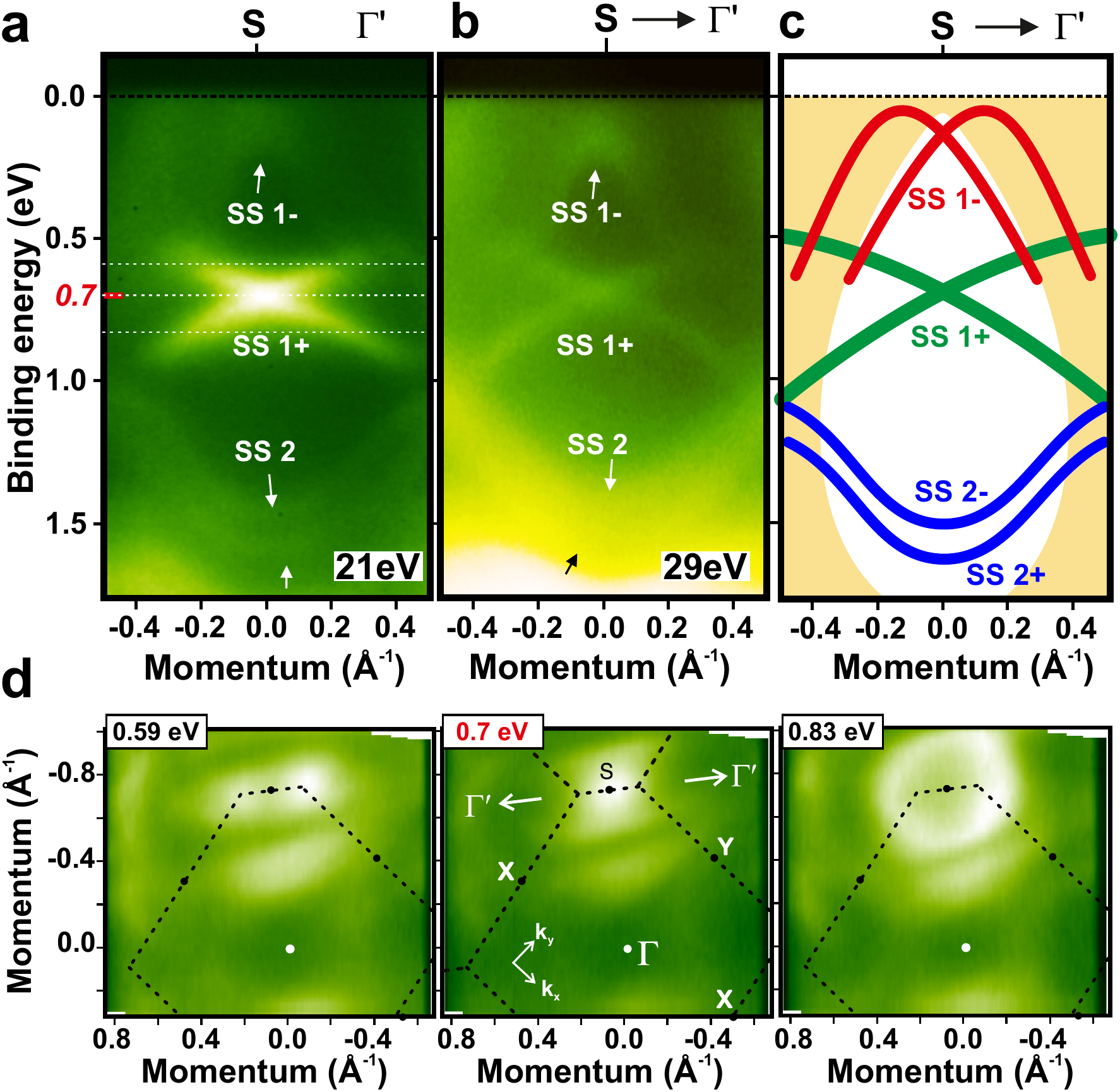}
\caption{\label{mFig2}(a) and (b) Intensity maps of the energy and $k$-resolved surface band structure of BiPd measured along the $\text{S}-\Gamma^{\prime}$ direction with $\nu=21~\mathrm{eV}$ and $\nu=29~\mathrm{eV}$, respectively. (c) Schematic representation of the surface states. (d) Constant energy cuts obtained at 0.59, 0.7, and 0.83~eV, energies indicated as dashed horizontal lines in (a). Overlaid on the constant energy cuts is a schematic of the surface Brillouin zone.}
\end{figure}

In full agreement between experiment and theory, the spin-splitting of the surface state is substantially larger in the S-$\Gamma^\prime$ direction compared to the S-$\Gamma$ direction. Experimental data for the S-$\Gamma^\prime$ direction are shown in Figs.~\ref{mFig2}(a) and (b), taken at the same photon energies $\nu$ as Figs.~\ref{mFig1}(d) and (e), respectively. The two measurements show the states SS1+ and SS1-- with different intensities, but otherwise at the same energy and having the same dispersion, confirming that they are of two-dimensional character. The different intensities are likely due to final state effects. There are small differences in binding energies between experiment and calculation on the order of 100~meV. One likely source of this discrepancy is surface relaxation which is neglected in the calculation. Constant energy contours obtained at the energies around the Dirac point, shown in Fig.~\ref{mFig2}(d) for the energies labelled in Fig.~\ref{mFig2}(a), clearly reveal the two band maxima in the S-$\Gamma^\prime$ direction due to the strongly anisotropic Rashba splitting (see also Ref.~\onlinecite{supp}).

Data and calculations yield a further set of surface states at higher binding energies, which we label SS2+ and SS2-. As opposed to the hole-like SS1$\pm$ states, SS2$\pm$ have an electron-like dispersion. In the experiment, they are most clearly resolved with $\nu=21~\mathrm{eV}$ (Fig.~\ref{mFig2}(a)). They are located near the bottom of the directional band gap at the S-point and quickly develop into surface resonances when moving away from $S$.

\begin{figure}[htb]
\includegraphics[width=0.9\columnwidth]{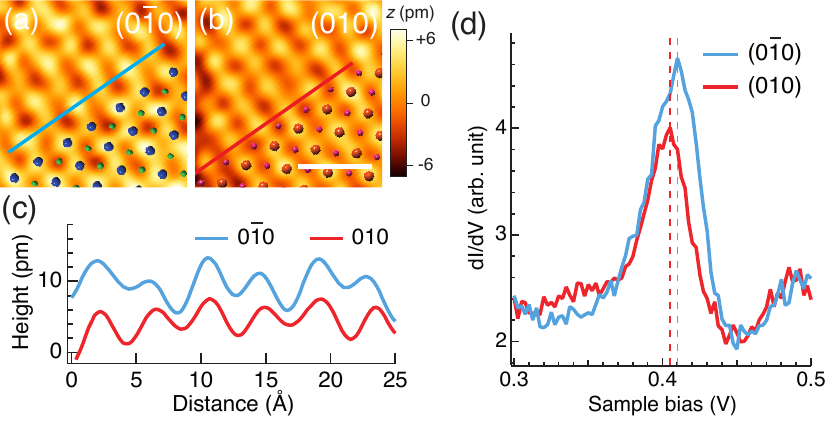}
\caption{\label{stmfig}(a) and (b) Topographies of the (0\=10) and (010) terminations respectively, obtained with the same tip. Blue/red spheres represent Bi atoms and green/purple Pd atoms in the top surface layer, compare fig.~\ref{mFig1}(a). (c) Linecuts of the two terminations, showing the different corrugations. Linecuts shifted horizontally for clarity. (d) $dI/dV$  spectra obtained on (0\=10) and (010) terminations. The surface state on the (0\=10) face is at a slightly larger energy than on (010) ($V_\mathrm s=0.5~\mathrm V$, $I_\mathrm s=2~\mathrm{nA}$).}
\end{figure}

Besides the surface states found at the $S$ points, the calculations reveal an additional pair of surfaces states at the $\Gamma$ point (labelled by SS0$\pm$ in Figs.~\ref{mFig1}(f) and (g)), which are in the unoccupied states and thus inaccessible to ARPES. For one termination, this state has been detected previously by STS~\cite{sunNatComm15}. While the Dirac-cone states at the $S$ point are present even if SOC is neglected, the Dirac state at the $\Gamma$ point appears within a gap opened up by SOC and arises as a consequence of an SOC-driven band inversion. This scenario is reminiscent of the topological insulator Bi$_2$Se$_3$~\cite{XiaNatPhys09}, indicating a possible topological origin. Here we show the signature of the surface state at $\Gamma$ for \emph{both} terminations from tunneling spectra, see Fig.~\ref{stmfig}. The terminations in the STM data have been identified from the surface corrugation (compare fig.~\ref{stmfig}(a-c)). Spectra of the surface state (fig.~\ref{stmfig}(d) show only a very small shift of $\sim 6~\mathrm{meV}$ between the two terminations, with the surface state showing up at larger energies on the termination which we identify as the (0\=10) surface.

We note that the band crossings of the Dirac states both at the $\Gamma$ and $S$ points are protected by time-reversal symmetry due to Kramers' theorem. Consistently for all surface states in the occupied states (SS1$\pm$, SS2$\pm$) those on the (010) surface occur at an energy at least 100~meV higher than on (0\=10), whereas the shift is very small and in the opposite direction for the surface state in the unoccupied states (SS0$\pm$).

\begin{figure}[t!]
\includegraphics[width=0.9\columnwidth]{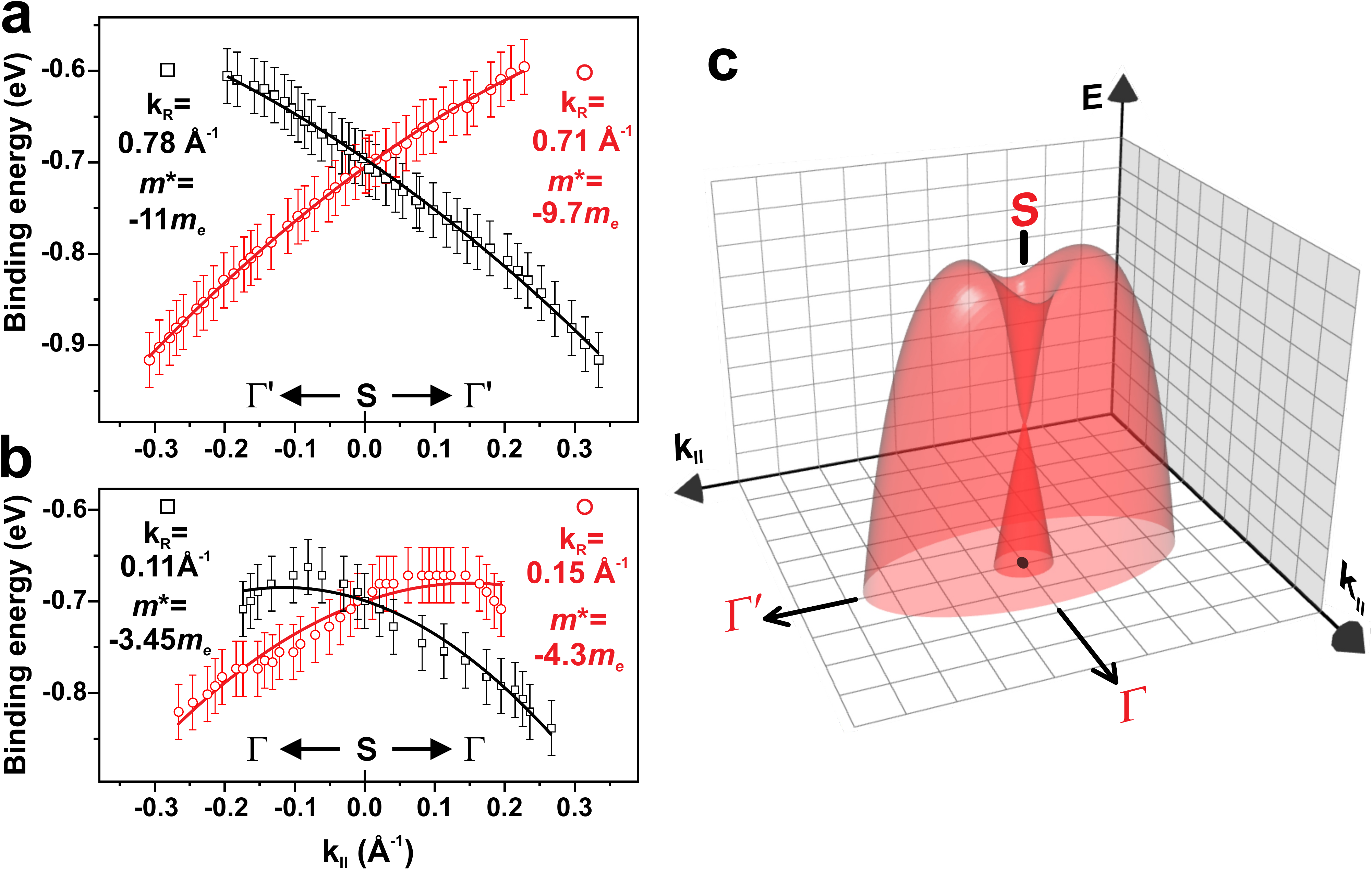}
\caption{\label{fig3fits}(a) and (b) Cuts in the S--$\Gamma^\prime$ and S--$\Gamma$ directions, respectively, from which the band structure parameters of
the SS1+ state have been determined. Solid lines show a fit of the Rashba model [Eq.~\eqref{Rashbamodel}] to the data. (c) Band structure of the Rashba-spin split surface state at the S-point (SS1+) as determined by fitting the Rashba model to the ARPES data. The Rashba spin splitting is highly anisotropic.}
\end{figure}

We have fitted the standard Rashba-Bychkov model~\cite{Bychkov1984} to cuts through the experimental band structure maps along the high-symmetry directions to extract the magnitude of spin splitting for the most prominent surface state, SS1+. The dispersion about the high-symmetry $S$ point is modeled as
\begin{equation}
E_{\pm}(k) = \frac{\hbar^2}{2 m^{\ast}} (\left |k\right|\pm k_\mathrm R)^2 + E_0,
\label{Rashbamodel}
\end{equation}
where $k$ denotes the momentum along the chosen direction in the surface BZ, $m^\ast$ is the effective mass, and $k_\mathrm R$ and $E_0$ denote the momentum offset and the energy of the band maxima, respectively.
We quantify the size of the Rashba splitting by the momentum offset $k_\mathrm R$ and the energy difference $E_\mathrm R = \hbar^2 k_\mathrm R^2 /( 2 m^{\ast})$ between the band maximum $E_0$ and the band crossing point.
The fits used to extract these parameters for SS1+ are shown in Figs.~\ref{fig3fits}(a) and (b) for the S--$\Gamma^\prime$ and S--$\Gamma$ directions, respectively. The Rashba momentum offset $k_\mathrm R$ and energy $E_\mathrm R$ along the S--$\Gamma^\prime$ direction in BiPd rank among the largest reported thus far, while both are significantly smaller in the S--$\Gamma$ direction. The results are summarized and compared with a selection of previously reported values in Table~\ref{tab:compare}. Despite the large momentum offset, the Rashba parameter $\alpha_\mathrm R=\hbar^2k_\mathrm R/m^\ast$ of BiPd is smaller than for the Bi/Ag(111) surface alloy due to the much larger effective mass of the surface states of BiPd.
Large Rashba splittings, leading to well-separated spin-split bands, may prove useful for applications involving the transport of spin rather than charge. Interesting, the Figure~\ref{fig3fits}(c) shows a three-dimensional representation of the dispersion of SS1+ near the S-point, highlighting the anisotropy in the Rashba spin-splitting.



\begin{table}[htb]
\caption{\label{tab:compare}Rashba momentum $k_\mathrm R$ in \AA$^{-1}$, Rashba energy $E_\mathrm R$ in meV, and Rashba parameter $\alpha_\mathrm R$ in eV$\cdot$\AA\ of materials with large Rashba-type band splitting.}
\begin{tabular}{lr@{.}lcr@{.}lc}\hline
Sample & \multicolumn{2}{c}{$k_\mathrm R$} & $E_\mathrm R$& \multicolumn{2}{c}{$\alpha_\mathrm R$} & Ref. \\ \hline\hline
Au(111) & 0&012 & 2.1 & 0&33 & \onlinecite{LaShell1996} \\
Bi(111) & 0&05 & 14 & 0&55 & \onlinecite{Koroteev2004} \\
Bi/Ag surface alloy & 0&13 & 200 & 3&05 & \onlinecite{astPRL07} \\
BiTeI & 0&052 & 100 & 3&8 & \onlinecite{IshizakaNatMat11} \\
BaNiS$_2$ & 0&2 & 150 & 0&26 & \onlinecite{Santos2016}\\
BiPd SS1+, S--$\Gamma$ &  0&13 & 17 & 0&25 & this work\\
BiPd SS1+, S--$\Gamma^{\prime}$ & 0&75 & 208 & 0 &55 & this work \\ \hline
\end{tabular}
\end{table}

Since BiPd is noncentrosymmetric and no symmetry element can transform a (010) surface into a (0\=10) surface, the shapes and energies of these surfaces' Dirac states can be quite different, and indeed this is what we observe.



Our data reveal a surprising richness of Dirac surface states on the (010)/(0\=10) surfaces of BiPd. Evidence for a surface state above $E_\mathrm F$ at $\Gamma$~\cite{sunNatComm15} and the observation of surface state SS1+ have been recently reported~\cite{Neupane1505} (although with a different assignment of the $S$ and $\Gamma$ points in the latter). From a detailed comparison of calculations, ARPES and STM data we can identify two distinct surface states below the Fermi level at the $S$ point and one at the $\Gamma$ point, on each surface. The data reveal signatures of surface states from opposite orientations of the cystallographic $b$-axis, which occur on opposite faces of an ideal crystal, implying twinning on the scale of the ARPES spot size. Macroscopic studies of the impact of the lack of inversion symmetry on the material properties may therefore need to detwin the material to yield information from a single domain. The overall consistency of our results with the previously published data confirms the high reproducibility of the properties of BiPd.


The Rashba splitting of the surface states at the S-point exhibits a strong anisotropy, suggesting strongly directionally-dependent SOC in the surface state. This strong directional dependence can be understood by comparison with the surface structure of BiPd: the $\Gamma$--S direction is along rows of Bi (or Pd) atoms, therefore electronic states propagating along this direction are only moderately exposed to the surface corrugation. Along the $\Gamma$--S$^\prime$ (or equivalently S--$\Gamma^\prime$) direction, rows of Bi and Pd atoms alternate, and electronic states with wavevectors along this direction are exposed much more strongly to the surface corrugation and hence to the surface electric fields which generate the spin splitting. The connection between surface corrugation and the spin-orbit splitting has been discussed previously in the context of the Bi/Ag(111) surface alloy \cite{gierz_structural_2010,bian_origin_2013}. In BiPd, the corrugation of the top-most layer is a direct consequence of the crystal structure of the bulk material, boosting the spin splitting of the surface states only in specific directions due to the anisotropy of the crystal structure.

In summary, through comparison of ARPES and STM experiments with band structure calculations, we have confirmed the presence of unconventional Dirac surface states in
noncentrosymmetric BiPd. The extremely large and anisotropic Rashba splitting in this system makes it an excellent candidate for future studies on the intricate spin texture of spin-split bands. Our results suggest a new way to engineer anisotropic spin textures and Rashba splittings of surface states by exploiting the low symmetry of the surface termination. The findings provide independent confirmation of the existence of twin boundaries in the material~\cite{sunNatComm15}, which may prove crucial to understanding its superconducting properties~\cite{Peets2016,Yan2016}.

\emph{Acknowledgments.---}
The authors thank Ed~Yelland for useful discussions. Funding from the MPG-UBC center and the Engineering and Physical Sciences Research Council (EP/I031014/1 and EP/L505079/1) are acknowledged.
This work was supported by the DFG within projects STA315/8-1 and BE5190/1-1. We also thank the staff at Bessy II of the Helmholtz-Zentrum Berlin for their assistance.

\bibliography{BiPd}

\begin{thebibliography}{53}%
\makeatletter
\providecommand \@ifxundefined [1]{%
 \@ifx{#1\undefined}
}%
\providecommand \@ifnum [1]{%
 \ifnum #1\expandafter \@firstoftwo
 \else \expandafter \@secondoftwo
 \fi
}%
\providecommand \@ifx [1]{%
 \ifx #1\expandafter \@firstoftwo
 \else \expandafter \@secondoftwo
 \fi
}%
\providecommand \natexlab [1]{#1}%
\providecommand \enquote  [1]{``#1''}%
\providecommand \bibnamefont  [1]{#1}%
\providecommand \bibfnamefont [1]{#1}%
\providecommand \citenamefont [1]{#1}%
\providecommand \href@noop [0]{\@secondoftwo}%
\providecommand \href [0]{\begingroup \@sanitize@url \@href}%
\providecommand \@href[1]{\@@startlink{#1}\@@href}%
\providecommand \@@href[1]{\endgroup#1\@@endlink}%
\providecommand \@sanitize@url [0]{\catcode `\\12\catcode `\$12\catcode
  `\&12\catcode `\#12\catcode `\^12\catcode `\_12\catcode `\%12\relax}%
\providecommand \@@startlink[1]{}%
\providecommand \@@endlink[0]{}%
\providecommand \url  [0]{\begingroup\@sanitize@url \@url }%
\providecommand \@url [1]{\endgroup\@href {#1}{\urlprefix }}%
\providecommand \urlprefix  [0]{URL }%
\providecommand \Eprint [0]{\href }%
\providecommand \doibase [0]{http://dx.doi.org/}%
\providecommand \selectlanguage [0]{\@gobble}%
\providecommand \bibinfo  [0]{\@secondoftwo}%
\providecommand \bibfield  [0]{\@secondoftwo}%
\providecommand \translation [1]{[#1]}%
\providecommand \BibitemOpen [0]{}%
\providecommand \bibitemStop [0]{}%
\providecommand \bibitemNoStop [0]{.\EOS\space}%
\providecommand \EOS [0]{\spacefactor3000\relax}%
\providecommand \BibitemShut  [1]{\csname bibitem#1\endcsname}%
\let\auto@bib@innerbib\@empty
\bibitem [{\citenamefont {Hasan}\ and\ \citenamefont
  {Kane}(2010)}]{hasanKaneRMP}%
  \BibitemOpen
  \bibfield  {author} {\bibinfo {author} {\bibfnamefont {M.~Z.}\ \bibnamefont
  {Hasan}}\ and\ \bibinfo {author} {\bibfnamefont {C.~L.}\ \bibnamefont
  {Kane}},\ }\href {\doibase 10.1103/RevModPhys.82.3045} {\bibfield  {journal}
  {\bibinfo  {journal} {Rev. Mod. Phys.}\ }\textbf {\bibinfo {volume} {82}},\
  \bibinfo {pages} {3045} (\bibinfo {year} {2010})}\BibitemShut {NoStop}%
\bibitem [{\citenamefont {Qi}\ and\ \citenamefont {Zhang}(2011)}]{qiZhangRMP}%
  \BibitemOpen
  \bibfield  {author} {\bibinfo {author} {\bibfnamefont {X.-L.}\ \bibnamefont
  {Qi}}\ and\ \bibinfo {author} {\bibfnamefont {S.-C.}\ \bibnamefont {Zhang}},\
  }\href {\doibase 10.1103/RevModPhys.83.1057} {\bibfield  {journal} {\bibinfo
  {journal} {Rev. Mod. Phys.}\ }\textbf {\bibinfo {volume} {83}},\ \bibinfo
  {pages} {1057} (\bibinfo {year} {2011})}\BibitemShut {NoStop}%
\bibitem [{\citenamefont {Hasan}\ and\ \citenamefont
  {Moore}(2011)}]{hasanMooreAnnRev}%
  \BibitemOpen
  \bibfield  {author} {\bibinfo {author} {\bibfnamefont {M.~Z.}\ \bibnamefont
  {Hasan}}\ and\ \bibinfo {author} {\bibfnamefont {J.~E.}\ \bibnamefont
  {Moore}},\ }\href {\doibase 10.1146/annurev-conmatphys-062910-140432}
  {\bibfield  {journal} {\bibinfo  {journal} {Ann. Rev. Cond. Matt. Phys.}\
  }\textbf {\bibinfo {volume} {2}},\ \bibinfo {pages} {55} (\bibinfo {year}
  {2011})}\BibitemShut {NoStop}%
\bibitem [{\citenamefont {Hsieh}\ \emph
  {et~al.}(2009{\natexlab{a}})\citenamefont {Hsieh}, \citenamefont {Xia},
  \citenamefont {Wray}, \citenamefont {Qian}, \citenamefont {Pal},
  \citenamefont {Dil}, \citenamefont {Osterwalder}, \citenamefont {Meier},
  \citenamefont {Bihlmayer}, \citenamefont {Kane}, \citenamefont {Hor},
  \citenamefont {Cava},\ and\ \citenamefont {Hasan}}]{D.Hsieh13022009}%
  \BibitemOpen
  \bibfield  {author} {\bibinfo {author} {\bibfnamefont {D.}~\bibnamefont
  {Hsieh}}, \bibinfo {author} {\bibfnamefont {Y.}~\bibnamefont {Xia}}, \bibinfo
  {author} {\bibfnamefont {L.}~\bibnamefont {Wray}}, \bibinfo {author}
  {\bibfnamefont {D.}~\bibnamefont {Qian}}, \bibinfo {author} {\bibfnamefont
  {A.}~\bibnamefont {Pal}}, \bibinfo {author} {\bibfnamefont {J.~H.}\
  \bibnamefont {Dil}}, \bibinfo {author} {\bibfnamefont {J.}~\bibnamefont
  {Osterwalder}}, \bibinfo {author} {\bibfnamefont {F.}~\bibnamefont {Meier}},
  \bibinfo {author} {\bibfnamefont {G.}~\bibnamefont {Bihlmayer}}, \bibinfo
  {author} {\bibfnamefont {C.~L.}\ \bibnamefont {Kane}}, \bibinfo {author}
  {\bibfnamefont {Y.~S.}\ \bibnamefont {Hor}}, \bibinfo {author} {\bibfnamefont
  {R.~J.}\ \bibnamefont {Cava}}, \ and\ \bibinfo {author} {\bibfnamefont
  {M.~Z.}\ \bibnamefont {Hasan}},\ }\href {\doibase 10.1126/science.1167733}
  {\bibfield  {journal} {\bibinfo  {journal} {Science}\ }\textbf {\bibinfo
  {volume} {323}},\ \bibinfo {pages} {919} (\bibinfo {year}
  {2009}{\natexlab{a}})}\BibitemShut {NoStop}%
\bibitem [{\citenamefont {Xia}\ \emph {et~al.}(2009)\citenamefont {Xia},
  \citenamefont {Qian}, \citenamefont {Hsieh}, \citenamefont {Wray},
  \citenamefont {Pal}, \citenamefont {Lin}, \citenamefont {Bansil},
  \citenamefont {Grauer}, \citenamefont {Hor}, \citenamefont {Cava},\ and\
  \citenamefont {Hasan}}]{XiaNatPhys09}%
  \BibitemOpen
  \bibfield  {author} {\bibinfo {author} {\bibfnamefont {Y.}~\bibnamefont
  {Xia}}, \bibinfo {author} {\bibfnamefont {D.}~\bibnamefont {Qian}}, \bibinfo
  {author} {\bibfnamefont {D.}~\bibnamefont {Hsieh}}, \bibinfo {author}
  {\bibfnamefont {L.}~\bibnamefont {Wray}}, \bibinfo {author} {\bibfnamefont
  {A.}~\bibnamefont {Pal}}, \bibinfo {author} {\bibfnamefont {H.}~\bibnamefont
  {Lin}}, \bibinfo {author} {\bibfnamefont {A.}~\bibnamefont {Bansil}},
  \bibinfo {author} {\bibfnamefont {D.}~\bibnamefont {Grauer}}, \bibinfo
  {author} {\bibfnamefont {Y.~S.}\ \bibnamefont {Hor}}, \bibinfo {author}
  {\bibfnamefont {R.~J.}\ \bibnamefont {Cava}}, \ and\ \bibinfo {author}
  {\bibfnamefont {M.~Z.}\ \bibnamefont {Hasan}},\ }\href {\doibase
  10.1038/nphys1274} {\bibfield  {journal} {\bibinfo  {journal} {Nat Phys}\
  }\textbf {\bibinfo {volume} {5}},\ \bibinfo {pages} {398} (\bibinfo {year}
  {2009})}\BibitemShut {NoStop}%
\bibitem [{\citenamefont {Seo}\ \emph {et~al.}(2010)\citenamefont {Seo},
  \citenamefont {Roushan}, \citenamefont {Beidenkopf}, \citenamefont {Hor},
  \citenamefont {Cava},\ and\ \citenamefont {Yazdani}}]{seoYazdaniNat10}%
  \BibitemOpen
  \bibfield  {author} {\bibinfo {author} {\bibfnamefont {J.}~\bibnamefont
  {Seo}}, \bibinfo {author} {\bibfnamefont {P.}~\bibnamefont {Roushan}},
  \bibinfo {author} {\bibfnamefont {H.}~\bibnamefont {Beidenkopf}}, \bibinfo
  {author} {\bibfnamefont {Y.~S.}\ \bibnamefont {Hor}}, \bibinfo {author}
  {\bibfnamefont {R.~J.}\ \bibnamefont {Cava}}, \ and\ \bibinfo {author}
  {\bibfnamefont {A.}~\bibnamefont {Yazdani}},\ }\href {\doibase
  10.1038/nature09189} {\bibfield  {journal} {\bibinfo  {journal} {Nature}\
  }\textbf {\bibinfo {volume} {466}},\ \bibinfo {pages} {343} (\bibinfo {year}
  {2010})}\BibitemShut {NoStop}%
\bibitem [{\citenamefont {Rashba}(1960)}]{rashba1960}%
  \BibitemOpen
  \bibfield  {author} {\bibinfo {author} {\bibfnamefont {E.~I.}\ \bibnamefont
  {Rashba}},\ }\href@noop {} {\bibfield  {journal} {\bibinfo  {journal} {Sov.
  Phys. Solid State}\ }\textbf {\bibinfo {volume} {2}},\ \bibinfo {pages}
  {1109} (\bibinfo {year} {1960})}\BibitemShut {NoStop}%
\bibitem [{\citenamefont {Ast}\ \emph {et~al.}(2007)\citenamefont {Ast},
  \citenamefont {Henk}, \citenamefont {Ernst}, \citenamefont {Moreschini},
  \citenamefont {Falub}, \citenamefont {Pacil\'e}, \citenamefont {Bruno},
  \citenamefont {Kern},\ and\ \citenamefont {Grioni}}]{astPRL07}%
  \BibitemOpen
  \bibfield  {author} {\bibinfo {author} {\bibfnamefont {C.~R.}\ \bibnamefont
  {Ast}}, \bibinfo {author} {\bibfnamefont {J.}~\bibnamefont {Henk}}, \bibinfo
  {author} {\bibfnamefont {A.}~\bibnamefont {Ernst}}, \bibinfo {author}
  {\bibfnamefont {L.}~\bibnamefont {Moreschini}}, \bibinfo {author}
  {\bibfnamefont {M.~C.}\ \bibnamefont {Falub}}, \bibinfo {author}
  {\bibfnamefont {D.}~\bibnamefont {Pacil\'e}}, \bibinfo {author}
  {\bibfnamefont {P.}~\bibnamefont {Bruno}}, \bibinfo {author} {\bibfnamefont
  {K.}~\bibnamefont {Kern}}, \ and\ \bibinfo {author} {\bibfnamefont
  {M.}~\bibnamefont {Grioni}},\ }\href {\doibase 10.1103/PhysRevLett.98.186807}
  {\bibfield  {journal} {\bibinfo  {journal} {Phys. Rev. Lett.}\ }\textbf
  {\bibinfo {volume} {98}},\ \bibinfo {pages} {186807} (\bibinfo {year}
  {2007})}\BibitemShut {NoStop}%
\bibitem [{\citenamefont {Crepaldi}\ \emph {et~al.}(2012)\citenamefont
  {Crepaldi}, \citenamefont {Moreschini}, \citenamefont {Aut\`es},
  \citenamefont {Tournier-Colletta}, \citenamefont {Moser}, \citenamefont
  {Virk}, \citenamefont {Berger}, \citenamefont {Bugnon}, \citenamefont
  {Chang}, \citenamefont {Kern}, \citenamefont {Bostwick}, \citenamefont
  {Rotenberg}, \citenamefont {Yazyev},\ and\ \citenamefont
  {Grioni}}]{crepaldiGrioniPRL12}%
  \BibitemOpen
  \bibfield  {author} {\bibinfo {author} {\bibfnamefont {A.}~\bibnamefont
  {Crepaldi}}, \bibinfo {author} {\bibfnamefont {L.}~\bibnamefont
  {Moreschini}}, \bibinfo {author} {\bibfnamefont {G.}~\bibnamefont {Aut\`es}},
  \bibinfo {author} {\bibfnamefont {C.}~\bibnamefont {Tournier-Colletta}},
  \bibinfo {author} {\bibfnamefont {S.}~\bibnamefont {Moser}}, \bibinfo
  {author} {\bibfnamefont {N.}~\bibnamefont {Virk}}, \bibinfo {author}
  {\bibfnamefont {H.}~\bibnamefont {Berger}}, \bibinfo {author} {\bibfnamefont
  {P.}~\bibnamefont {Bugnon}}, \bibinfo {author} {\bibfnamefont {Y.~J.}\
  \bibnamefont {Chang}}, \bibinfo {author} {\bibfnamefont {K.}~\bibnamefont
  {Kern}}, \bibinfo {author} {\bibfnamefont {A.}~\bibnamefont {Bostwick}},
  \bibinfo {author} {\bibfnamefont {E.}~\bibnamefont {Rotenberg}}, \bibinfo
  {author} {\bibfnamefont {O.~V.}\ \bibnamefont {Yazyev}}, \ and\ \bibinfo
  {author} {\bibfnamefont {M.}~\bibnamefont {Grioni}},\ }\href {\doibase
  10.1103/PhysRevLett.109.096803} {\bibfield  {journal} {\bibinfo  {journal}
  {Phys. Rev. Lett.}\ }\textbf {\bibinfo {volume} {109}},\ \bibinfo {pages}
  {096803} (\bibinfo {year} {2012})}\BibitemShut {NoStop}%
\bibitem [{\citenamefont {Bahramy}\ \emph {et~al.}(2012)\citenamefont
  {Bahramy}, \citenamefont {Yang}, \citenamefont {Arita},\ and\ \citenamefont
  {Nagaosa}}]{bahramyNatComm12}%
  \BibitemOpen
  \bibfield  {author} {\bibinfo {author} {\bibfnamefont {M.}~\bibnamefont
  {Bahramy}}, \bibinfo {author} {\bibfnamefont {B.-J.}\ \bibnamefont {Yang}},
  \bibinfo {author} {\bibfnamefont {R.}~\bibnamefont {Arita}}, \ and\ \bibinfo
  {author} {\bibfnamefont {N.}~\bibnamefont {Nagaosa}},\ }\href {\doibase
  10.1038/ncomms1679} {\bibfield  {journal} {\bibinfo  {journal} {Nat.
  Commun.}\ }\textbf {\bibinfo {volume} {3}},\ \bibinfo {pages} {679} (\bibinfo
  {year} {2012})}\BibitemShut {NoStop}%
\bibitem [{\citenamefont {Hsieh}\ \emph
  {et~al.}(2009{\natexlab{b}})\citenamefont {Hsieh}, \citenamefont {Xia},
  \citenamefont {Qian}, \citenamefont {Wray}, \citenamefont {Dil},
  \citenamefont {Meier}, \citenamefont {Osterwalder}, \citenamefont {Patthey},
  \citenamefont {Checkelsky}, \citenamefont {Ong}, \citenamefont {Fedorov},
  \citenamefont {Lin}, \citenamefont {Bansil}, \citenamefont {Grauer},
  \citenamefont {Hor}, \citenamefont {Cava},\ and\ \citenamefont
  {Hasan}}]{hsiehHasanNature09}%
  \BibitemOpen
  \bibfield  {author} {\bibinfo {author} {\bibfnamefont {D.}~\bibnamefont
  {Hsieh}}, \bibinfo {author} {\bibfnamefont {Y.}~\bibnamefont {Xia}}, \bibinfo
  {author} {\bibfnamefont {D.}~\bibnamefont {Qian}}, \bibinfo {author}
  {\bibfnamefont {L.}~\bibnamefont {Wray}}, \bibinfo {author} {\bibfnamefont
  {J.~H.}\ \bibnamefont {Dil}}, \bibinfo {author} {\bibfnamefont
  {F.}~\bibnamefont {Meier}}, \bibinfo {author} {\bibfnamefont
  {J.}~\bibnamefont {Osterwalder}}, \bibinfo {author} {\bibfnamefont
  {L.}~\bibnamefont {Patthey}}, \bibinfo {author} {\bibfnamefont {J.~G.}\
  \bibnamefont {Checkelsky}}, \bibinfo {author} {\bibfnamefont {N.~P.}\
  \bibnamefont {Ong}}, \bibinfo {author} {\bibfnamefont {A.~V.}\ \bibnamefont
  {Fedorov}}, \bibinfo {author} {\bibfnamefont {H.}~\bibnamefont {Lin}},
  \bibinfo {author} {\bibfnamefont {A.}~\bibnamefont {Bansil}}, \bibinfo
  {author} {\bibfnamefont {D.}~\bibnamefont {Grauer}}, \bibinfo {author}
  {\bibfnamefont {Y.~S.}\ \bibnamefont {Hor}}, \bibinfo {author} {\bibfnamefont
  {R.~J.}\ \bibnamefont {Cava}}, \ and\ \bibinfo {author} {\bibfnamefont
  {M.~Z.}\ \bibnamefont {Hasan}},\ }\href {\doibase 10.1038/nature08234}
  {\bibfield  {journal} {\bibinfo  {journal} {Nature}\ }\textbf {\bibinfo
  {volume} {460}},\ \bibinfo {pages} {1101} (\bibinfo {year}
  {2009}{\natexlab{b}})}\BibitemShut {NoStop}%
\bibitem [{\citenamefont {Ishizaka}\ \emph {et~al.}(2011)\citenamefont
  {Ishizaka}, \citenamefont {Bahramy}, \citenamefont {Murakawa}, \citenamefont
  {Sakano}, \citenamefont {Shimojima}, \citenamefont {Sonobe}, \citenamefont
  {Koizumi}, \citenamefont {Shin}, \citenamefont {Miyahara}, \citenamefont
  {Kimura}, \citenamefont {Miyamoto}, \citenamefont {Okuda}, \citenamefont
  {Namatame}, \citenamefont {Taniguchi}, \citenamefont {Arita}, \citenamefont
  {Nagaosa}, \citenamefont {Kobayashi}, \citenamefont {Murakami}, \citenamefont
  {Kumai}, \citenamefont {Kaneko}, \citenamefont {Onose},\ and\ \citenamefont
  {Tokura}}]{IshizakaNatMat11}%
  \BibitemOpen
  \bibfield  {author} {\bibinfo {author} {\bibfnamefont {K.}~\bibnamefont
  {Ishizaka}}, \bibinfo {author} {\bibfnamefont {M.~S.}\ \bibnamefont
  {Bahramy}}, \bibinfo {author} {\bibfnamefont {H.}~\bibnamefont {Murakawa}},
  \bibinfo {author} {\bibfnamefont {M.}~\bibnamefont {Sakano}}, \bibinfo
  {author} {\bibfnamefont {T.}~\bibnamefont {Shimojima}}, \bibinfo {author}
  {\bibfnamefont {T.}~\bibnamefont {Sonobe}}, \bibinfo {author} {\bibfnamefont
  {K.}~\bibnamefont {Koizumi}}, \bibinfo {author} {\bibfnamefont
  {S.}~\bibnamefont {Shin}}, \bibinfo {author} {\bibfnamefont {H.}~\bibnamefont
  {Miyahara}}, \bibinfo {author} {\bibfnamefont {A.}~\bibnamefont {Kimura}},
  \bibinfo {author} {\bibfnamefont {K.}~\bibnamefont {Miyamoto}}, \bibinfo
  {author} {\bibfnamefont {T.}~\bibnamefont {Okuda}}, \bibinfo {author}
  {\bibfnamefont {H.}~\bibnamefont {Namatame}}, \bibinfo {author}
  {\bibfnamefont {M.}~\bibnamefont {Taniguchi}}, \bibinfo {author}
  {\bibfnamefont {R.}~\bibnamefont {Arita}}, \bibinfo {author} {\bibfnamefont
  {N.}~\bibnamefont {Nagaosa}}, \bibinfo {author} {\bibfnamefont
  {K.}~\bibnamefont {Kobayashi}}, \bibinfo {author} {\bibfnamefont
  {Y.}~\bibnamefont {Murakami}}, \bibinfo {author} {\bibfnamefont
  {R.}~\bibnamefont {Kumai}}, \bibinfo {author} {\bibfnamefont
  {Y.}~\bibnamefont {Kaneko}}, \bibinfo {author} {\bibfnamefont
  {Y.}~\bibnamefont {Onose}}, \ and\ \bibinfo {author} {\bibfnamefont
  {Y.}~\bibnamefont {Tokura}},\ }\href {\doibase 10.1038/nmat3051} {\bibfield
  {journal} {\bibinfo  {journal} {Nat. Mater.}\ }\textbf {\bibinfo {volume}
  {10}},\ \bibinfo {pages} {521} (\bibinfo {year} {2011})}\BibitemShut
  {NoStop}%
\bibitem [{\citenamefont {Wang}\ \emph {et~al.}(2011)\citenamefont {Wang},
  \citenamefont {Hsieh}, \citenamefont {Pilon}, \citenamefont {Fu},
  \citenamefont {Gardner}, \citenamefont {Lee},\ and\ \citenamefont
  {Gedik}}]{wangGedikPRL11}%
  \BibitemOpen
  \bibfield  {author} {\bibinfo {author} {\bibfnamefont {Y.~H.}\ \bibnamefont
  {Wang}}, \bibinfo {author} {\bibfnamefont {D.}~\bibnamefont {Hsieh}},
  \bibinfo {author} {\bibfnamefont {D.}~\bibnamefont {Pilon}}, \bibinfo
  {author} {\bibfnamefont {L.}~\bibnamefont {Fu}}, \bibinfo {author}
  {\bibfnamefont {D.~R.}\ \bibnamefont {Gardner}}, \bibinfo {author}
  {\bibfnamefont {Y.~S.}\ \bibnamefont {Lee}}, \ and\ \bibinfo {author}
  {\bibfnamefont {N.}~\bibnamefont {Gedik}},\ }\href {\doibase
  10.1103/PhysRevLett.107.207602} {\bibfield  {journal} {\bibinfo  {journal}
  {Phys. Rev. Lett.}\ }\textbf {\bibinfo {volume} {107}},\ \bibinfo {pages}
  {207602} (\bibinfo {year} {2011})}\BibitemShut {NoStop}%
\bibitem [{\citenamefont {Bauer}\ and\ \citenamefont
  {Sigrist}(2012)}]{bauerSigristBook}%
  \BibitemOpen
  \bibfield  {author} {\bibinfo {author} {\bibfnamefont {E.}~\bibnamefont
  {Bauer}}\ and\ \bibinfo {author} {\bibfnamefont {M.}~\bibnamefont
  {Sigrist}},\ }\href {\doibase 10.1007/978-3-642-24624-1} {\emph {\bibinfo
  {title} {Non-Centrosymmetric Superconductors: Introduction and Overview}}},\
  \bibinfo {series} {Lecture Notes in Physics}, Vol.\ \bibinfo {volume} {847}\
  (\bibinfo  {publisher} {Springer Berlin},\ \bibinfo {year} {2012})\ pp.\
  \bibinfo {pages} {1--357}\BibitemShut {NoStop}%
\bibitem [{\citenamefont {Bauer}\ \emph {et~al.}(2004)\citenamefont {Bauer},
  \citenamefont {Hilscher}, \citenamefont {Michor}, \citenamefont {Paul},
  \citenamefont {Scheidt}, \citenamefont {Gribanov}, \citenamefont {Seropegin},
  \citenamefont {No{\"e}l}, \citenamefont {Sigrist},\ and\ \citenamefont
  {Rogl}}]{bauer2004heavy}%
  \BibitemOpen
  \bibfield  {author} {\bibinfo {author} {\bibfnamefont {E.}~\bibnamefont
  {Bauer}}, \bibinfo {author} {\bibfnamefont {G.}~\bibnamefont {Hilscher}},
  \bibinfo {author} {\bibfnamefont {H.}~\bibnamefont {Michor}}, \bibinfo
  {author} {\bibfnamefont {C.}~\bibnamefont {Paul}}, \bibinfo {author}
  {\bibfnamefont {E.~W.}\ \bibnamefont {Scheidt}}, \bibinfo {author}
  {\bibfnamefont {A.}~\bibnamefont {Gribanov}}, \bibinfo {author}
  {\bibfnamefont {Y.}~\bibnamefont {Seropegin}}, \bibinfo {author}
  {\bibfnamefont {H.}~\bibnamefont {No{\"e}l}}, \bibinfo {author}
  {\bibfnamefont {M.}~\bibnamefont {Sigrist}}, \ and\ \bibinfo {author}
  {\bibfnamefont {P.}~\bibnamefont {Rogl}},\ }\href {\doibase
  10.1103/PhysRevLett.92.027003} {\bibfield  {journal} {\bibinfo  {journal}
  {Phys. Rev. Lett.}\ }\textbf {\bibinfo {volume} {92}},\ \bibinfo {pages}
  {027003} (\bibinfo {year} {2004})}\BibitemShut {NoStop}%
\bibitem [{\citenamefont {Schnyder}\ \emph {et~al.}(2008)\citenamefont
  {Schnyder}, \citenamefont {Ryu}, \citenamefont {Furusaki},\ and\
  \citenamefont {Ludwig}}]{Schnyder2008gfback}%
  \BibitemOpen
  \bibfield  {author} {\bibinfo {author} {\bibfnamefont {A.~P.}\ \bibnamefont
  {Schnyder}}, \bibinfo {author} {\bibfnamefont {S.}~\bibnamefont {Ryu}},
  \bibinfo {author} {\bibfnamefont {A.}~\bibnamefont {Furusaki}}, \ and\
  \bibinfo {author} {\bibfnamefont {A.~W.~W.}\ \bibnamefont {Ludwig}},\ }\href
  {\doibase 10.1103/PhysRevB.78.195125} {\bibfield  {journal} {\bibinfo
  {journal} {Phys. Rev. B}\ }\textbf {\bibinfo {volume} {78}},\ \bibinfo
  {pages} {195125} (\bibinfo {year} {2008})}\BibitemShut {NoStop}%
\bibitem [{\citenamefont {B\'eri}(2010)}]{beriPRB10}%
  \BibitemOpen
  \bibfield  {author} {\bibinfo {author} {\bibfnamefont {B.}~\bibnamefont
  {B\'eri}},\ }\href {\doibase 10.1103/PhysRevB.81.134515} {\bibfield
  {journal} {\bibinfo  {journal} {Phys. Rev. B}\ }\textbf {\bibinfo {volume}
  {81}},\ \bibinfo {pages} {134515} (\bibinfo {year} {2010})}\BibitemShut
  {NoStop}%
\bibitem [{\citenamefont {Schnyder}\ and\ \citenamefont
  {Ryu}(2011)}]{Andreas_flat}%
  \BibitemOpen
  \bibfield  {author} {\bibinfo {author} {\bibfnamefont {A.~P.}\ \bibnamefont
  {Schnyder}}\ and\ \bibinfo {author} {\bibfnamefont {S.}~\bibnamefont {Ryu}},\
  }\href {\doibase 10.1103/PhysRevB.84.060504} {\bibfield  {journal} {\bibinfo
  {journal} {Phys. Rev. B}\ }\textbf {\bibinfo {volume} {84}},\ \bibinfo
  {pages} {060504} (\bibinfo {year} {2011})}\BibitemShut {NoStop}%
\bibitem [{\citenamefont {Sato}\ and\ \citenamefont
  {Fujimoto}(2009)}]{satoFujimotoPRB09}%
  \BibitemOpen
  \bibfield  {author} {\bibinfo {author} {\bibfnamefont {M.}~\bibnamefont
  {Sato}}\ and\ \bibinfo {author} {\bibfnamefont {S.}~\bibnamefont
  {Fujimoto}},\ }\href {\doibase 10.1103/PhysRevB.79.094504} {\bibfield
  {journal} {\bibinfo  {journal} {Phys. Rev. B}\ }\textbf {\bibinfo {volume}
  {79}},\ \bibinfo {pages} {094504} (\bibinfo {year} {2009})}\BibitemShut
  {NoStop}%
\bibitem [{\citenamefont {Kheiker}\ \emph {et~al.}(1953)\citenamefont
  {Kheiker}, \citenamefont {Zhdanov},\ and\ \citenamefont
  {Zhuravlev}}]{kheiker1953x}%
  \BibitemOpen
  \bibfield  {author} {\bibinfo {author} {\bibfnamefont {D.~M.}\ \bibnamefont
  {Kheiker}}, \bibinfo {author} {\bibfnamefont {G.~S.}\ \bibnamefont
  {Zhdanov}}, \ and\ \bibinfo {author} {\bibfnamefont {N.~N.}\ \bibnamefont
  {Zhuravlev}},\ }\href@noop {} {\bibfield  {journal} {\bibinfo  {journal} {Zh.
  Eksp. Teor. Fiz.}\ }\textbf {\bibinfo {volume} {25}},\ \bibinfo {pages} {621}
  (\bibinfo {year} {1953})}\BibitemShut {NoStop}%
\bibitem [{\citenamefont {Zhuravlev}(1957)}]{Zhuravlev1957}%
  \BibitemOpen
  \bibfield  {author} {\bibinfo {author} {\bibfnamefont {N.}~\bibnamefont
  {Zhuravlev}},\ }\href@noop {} {\bibfield  {journal} {\bibinfo  {journal} {Zh.
  Eksp. Teor. Fiz.}\ }\textbf {\bibinfo {volume} {5}},\ \bibinfo {pages} {1064}
  (\bibinfo {year} {1957})}\BibitemShut {NoStop}%
\bibitem [{\citenamefont {Bhatt}\ and\ \citenamefont
  {Schubert}(1980)}]{bhatt1980kristallstruktur}%
  \BibitemOpen
  \bibfield  {author} {\bibinfo {author} {\bibfnamefont {Y.}~\bibnamefont
  {Bhatt}}\ and\ \bibinfo {author} {\bibfnamefont {K.}~\bibnamefont
  {Schubert}},\ }\href {\doibase 10.1016/0022-5088(80)90285-4} {\bibfield
  {journal} {\bibinfo  {journal} {Journal of the Less Common Metals}\ }\textbf
  {\bibinfo {volume} {70}},\ \bibinfo {pages} {P39} (\bibinfo {year}
  {1980})}\BibitemShut {NoStop}%
\bibitem [{\citenamefont {Bhatt}\ and\ \citenamefont
  {Schubert}(1979)}]{Bhatt1979P17}%
  \BibitemOpen
  \bibfield  {author} {\bibinfo {author} {\bibfnamefont {Y.}~\bibnamefont
  {Bhatt}}\ and\ \bibinfo {author} {\bibfnamefont {K.}~\bibnamefont
  {Schubert}},\ }\href {\doibase 10.1016/0022-5088(79)90184-X} {\bibfield
  {journal} {\bibinfo  {journal} {J. Less-Common Met.}\ }\textbf {\bibinfo
  {volume} {64}},\ \bibinfo {pages} {P17 } (\bibinfo {year}
  {1979})}\BibitemShut {NoStop}%
\bibitem [{\citenamefont {Ionov}\ \emph {et~al.}(1989)\citenamefont {Ionov},
  \citenamefont {Tomilin}, \citenamefont {Prozorovskii}, \citenamefont
  {Klimenko}, \citenamefont {Titov}, \citenamefont {Zhukov},\ and\
  \citenamefont {Fetisov}}]{Ionov1989}%
  \BibitemOpen
  \bibfield  {author} {\bibinfo {author} {\bibfnamefont {V.~M.}\ \bibnamefont
  {Ionov}}, \bibinfo {author} {\bibfnamefont {N.~A.}\ \bibnamefont {Tomilin}},
  \bibinfo {author} {\bibfnamefont {A.~E.}\ \bibnamefont {Prozorovskii}},
  \bibinfo {author} {\bibfnamefont {A.~N.}\ \bibnamefont {Klimenko}}, \bibinfo
  {author} {\bibfnamefont {Y.~V.}\ \bibnamefont {Titov}}, \bibinfo {author}
  {\bibfnamefont {S.~G.}\ \bibnamefont {Zhukov}}, \ and\ \bibinfo {author}
  {\bibfnamefont {G.~V.}\ \bibnamefont {Fetisov}},\ }\href@noop {} {\bibfield
  {journal} {\bibinfo  {journal} {Sov. Phys. Crystallogr.}\ }\textbf {\bibinfo
  {volume} {34}},\ \bibinfo {pages} {496} (\bibinfo {year} {1989})}\BibitemShut
  {NoStop}%
\bibitem [{\citenamefont {Alekseevskii}(1952)}]{Alekseevskii1952}%
  \BibitemOpen
  \bibfield  {author} {\bibinfo {author} {\bibfnamefont {N.~E.}\ \bibnamefont
  {Alekseevskii}},\ }\href@noop {} {\bibfield  {journal} {\bibinfo  {journal}
  {Zh.\ Eksp.\ Teor.\ Fiz.}\ }\textbf {\bibinfo {volume} {23}},\ \bibinfo
  {pages} {484} (\bibinfo {year} {1952})}\BibitemShut {NoStop}%
\bibitem [{\citenamefont {Joshi}\ \emph {et~al.}(2011)\citenamefont {Joshi},
  \citenamefont {Thamizhavel},\ and\ \citenamefont
  {Ramakrishnan}}]{joshi2011superconductivity}%
  \BibitemOpen
  \bibfield  {author} {\bibinfo {author} {\bibfnamefont {B.}~\bibnamefont
  {Joshi}}, \bibinfo {author} {\bibfnamefont {A.}~\bibnamefont {Thamizhavel}},
  \ and\ \bibinfo {author} {\bibfnamefont {S.}~\bibnamefont {Ramakrishnan}},\
  }\href {\doibase 10.1103/PhysRevB.84.064518} {\bibfield  {journal} {\bibinfo
  {journal} {Phys. Rev. B}\ }\textbf {\bibinfo {volume} {84}},\ \bibinfo
  {pages} {064518} (\bibinfo {year} {2011})}\BibitemShut {NoStop}%
\bibitem [{\citenamefont {Mondal}\ \emph {et~al.}(2012)\citenamefont {Mondal},
  \citenamefont {Joshi}, \citenamefont {Kumar}, \citenamefont {Kamlapure},
  \citenamefont {Ganguli}, \citenamefont {Thamizhavel}, \citenamefont {Mandal},
  \citenamefont {Ramakrishnan},\ and\ \citenamefont
  {Raychaudhuri}}]{mondal2012andreev}%
  \BibitemOpen
  \bibfield  {author} {\bibinfo {author} {\bibfnamefont {M.}~\bibnamefont
  {Mondal}}, \bibinfo {author} {\bibfnamefont {B.}~\bibnamefont {Joshi}},
  \bibinfo {author} {\bibfnamefont {S.}~\bibnamefont {Kumar}}, \bibinfo
  {author} {\bibfnamefont {A.}~\bibnamefont {Kamlapure}}, \bibinfo {author}
  {\bibfnamefont {S.~C.}\ \bibnamefont {Ganguli}}, \bibinfo {author}
  {\bibfnamefont {A.}~\bibnamefont {Thamizhavel}}, \bibinfo {author}
  {\bibfnamefont {S.~S.}\ \bibnamefont {Mandal}}, \bibinfo {author}
  {\bibfnamefont {S.}~\bibnamefont {Ramakrishnan}}, \ and\ \bibinfo {author}
  {\bibfnamefont {P.}~\bibnamefont {Raychaudhuri}},\ }\href {\doibase
  10.1103/PhysRevB.86.094520} {\bibfield  {journal} {\bibinfo  {journal} {Phys.
  Rev. B}\ }\textbf {\bibinfo {volume} {86}},\ \bibinfo {pages} {094520}
  (\bibinfo {year} {2012})}\BibitemShut {NoStop}%
\bibitem [{\citenamefont {Matano}\ \emph {et~al.}(2013)\citenamefont {Matano},
  \citenamefont {Maeda}, \citenamefont {Sawaoka}, \citenamefont {Muro},
  \citenamefont {Takabatake}, \citenamefont {Joshi}, \citenamefont
  {Ramakrishnan}, \citenamefont {Kawashima}, \citenamefont {Akimitsu},\ and\
  \citenamefont {Zheng}}]{matano2013nmr}%
  \BibitemOpen
  \bibfield  {author} {\bibinfo {author} {\bibfnamefont {K.}~\bibnamefont
  {Matano}}, \bibinfo {author} {\bibfnamefont {S.}~\bibnamefont {Maeda}},
  \bibinfo {author} {\bibfnamefont {H.}~\bibnamefont {Sawaoka}}, \bibinfo
  {author} {\bibfnamefont {Y.}~\bibnamefont {Muro}}, \bibinfo {author}
  {\bibfnamefont {T.}~\bibnamefont {Takabatake}}, \bibinfo {author}
  {\bibfnamefont {B.}~\bibnamefont {Joshi}}, \bibinfo {author} {\bibfnamefont
  {S.}~\bibnamefont {Ramakrishnan}}, \bibinfo {author} {\bibfnamefont
  {K.}~\bibnamefont {Kawashima}}, \bibinfo {author} {\bibfnamefont
  {J.}~\bibnamefont {Akimitsu}}, \ and\ \bibinfo {author} {\bibfnamefont
  {G.-q.}\ \bibnamefont {Zheng}},\ }\href {\doibase 10.7566/JPSJ.82.084711}
  {\bibfield  {journal} {\bibinfo  {journal} {J. Phys. Soc. Jpn.}\ }\textbf
  {\bibinfo {volume} {82}},\ \bibinfo {pages} {084711} (\bibinfo {year}
  {2013})}\BibitemShut {NoStop}%
\bibitem [{\citenamefont {Sun}\ \emph {et~al.}(2015)\citenamefont {Sun},
  \citenamefont {Enayat}, \citenamefont {Maldonado}, \citenamefont {Lithgow},
  \citenamefont {Yelland}, \citenamefont {Peets}, \citenamefont {Yaresko},
  \citenamefont {Schnyder},\ and\ \citenamefont {Wahl}}]{sunNatComm15}%
  \BibitemOpen
  \bibfield  {author} {\bibinfo {author} {\bibfnamefont {Z.}~\bibnamefont
  {Sun}}, \bibinfo {author} {\bibfnamefont {M.}~\bibnamefont {Enayat}},
  \bibinfo {author} {\bibfnamefont {A.}~\bibnamefont {Maldonado}}, \bibinfo
  {author} {\bibfnamefont {C.}~\bibnamefont {Lithgow}}, \bibinfo {author}
  {\bibfnamefont {E.}~\bibnamefont {Yelland}}, \bibinfo {author} {\bibfnamefont
  {D.~C.}\ \bibnamefont {Peets}}, \bibinfo {author} {\bibfnamefont
  {A.}~\bibnamefont {Yaresko}}, \bibinfo {author} {\bibfnamefont {A.~P.}\
  \bibnamefont {Schnyder}}, \ and\ \bibinfo {author} {\bibfnamefont
  {P.}~\bibnamefont {Wahl}},\ }\href {\doibase 10.1038/ncomms7633} {\bibfield
  {journal} {\bibinfo  {journal} {Nat Commun}\ }\textbf {\bibinfo {volume}
  {6}},\ \bibinfo {pages} {6633} (\bibinfo {year} {2015})}\BibitemShut
  {NoStop}%
\bibitem [{\citenamefont {Peets}\ \emph {et~al.}(2016)\citenamefont {Peets},
  \citenamefont {Maldonado}, \citenamefont {Enayat}, \citenamefont {Sun},
  \citenamefont {Wahl},\ and\ \citenamefont {Schnyder}}]{Peets2016}%
  \BibitemOpen
  \bibfield  {author} {\bibinfo {author} {\bibfnamefont {D.~C.}\ \bibnamefont
  {Peets}}, \bibinfo {author} {\bibfnamefont {A.}~\bibnamefont {Maldonado}},
  \bibinfo {author} {\bibfnamefont {M.}~\bibnamefont {Enayat}}, \bibinfo
  {author} {\bibfnamefont {Z.}~\bibnamefont {Sun}}, \bibinfo {author}
  {\bibfnamefont {P.}~\bibnamefont {Wahl}}, \ and\ \bibinfo {author}
  {\bibfnamefont {A.~P.}\ \bibnamefont {Schnyder}},\ }\href {\doibase
  10.1103/PhysRevB.93.174504} {\bibfield  {journal} {\bibinfo  {journal} {Phys.
  Rev. B}\ }\textbf {\bibinfo {volume} {93}},\ \bibinfo {pages} {174504}
  (\bibinfo {year} {2016})}\BibitemShut {NoStop}%
\bibitem [{\citenamefont {Sasaki}\ \emph {et~al.}(2011)\citenamefont {Sasaki},
  \citenamefont {Kriener}, \citenamefont {Segawa}, \citenamefont {Yada},
  \citenamefont {Tanaka}, \citenamefont {Sato},\ and\ \citenamefont
  {Ando}}]{sasakiAndoPRL11}%
  \BibitemOpen
  \bibfield  {author} {\bibinfo {author} {\bibfnamefont {S.}~\bibnamefont
  {Sasaki}}, \bibinfo {author} {\bibfnamefont {M.}~\bibnamefont {Kriener}},
  \bibinfo {author} {\bibfnamefont {K.}~\bibnamefont {Segawa}}, \bibinfo
  {author} {\bibfnamefont {K.}~\bibnamefont {Yada}}, \bibinfo {author}
  {\bibfnamefont {Y.}~\bibnamefont {Tanaka}}, \bibinfo {author} {\bibfnamefont
  {M.}~\bibnamefont {Sato}}, \ and\ \bibinfo {author} {\bibfnamefont
  {Y.}~\bibnamefont {Ando}},\ }\href {\doibase 10.1103/PhysRevLett.107.217001}
  {\bibfield  {journal} {\bibinfo  {journal} {Phys. Rev. Lett.}\ }\textbf
  {\bibinfo {volume} {107}},\ \bibinfo {pages} {217001} (\bibinfo {year}
  {2011})}\BibitemShut {NoStop}%
\bibitem [{\citenamefont {Levy}\ \emph {et~al.}(2013)\citenamefont {Levy},
  \citenamefont {Zhang}, \citenamefont {Ha}, \citenamefont {Sharifi},
  \citenamefont {Talin}, \citenamefont {Kuk},\ and\ \citenamefont
  {Stroscio}}]{levyStroscioPRL13}%
  \BibitemOpen
  \bibfield  {author} {\bibinfo {author} {\bibfnamefont {N.}~\bibnamefont
  {Levy}}, \bibinfo {author} {\bibfnamefont {T.}~\bibnamefont {Zhang}},
  \bibinfo {author} {\bibfnamefont {J.}~\bibnamefont {Ha}}, \bibinfo {author}
  {\bibfnamefont {F.}~\bibnamefont {Sharifi}}, \bibinfo {author} {\bibfnamefont
  {A.~A.}\ \bibnamefont {Talin}}, \bibinfo {author} {\bibfnamefont
  {Y.}~\bibnamefont {Kuk}}, \ and\ \bibinfo {author} {\bibfnamefont {J.~A.}\
  \bibnamefont {Stroscio}},\ }\href {\doibase 10.1103/PhysRevLett.110.117001}
  {\bibfield  {journal} {\bibinfo  {journal} {Phys. Rev. Lett.}\ }\textbf
  {\bibinfo {volume} {110}},\ \bibinfo {pages} {117001} (\bibinfo {year}
  {2013})}\BibitemShut {NoStop}%
\bibitem [{\citenamefont {Liu}\ \emph {et~al.}(2011)\citenamefont {Liu},
  \citenamefont {Lee}, \citenamefont {Kondo}, \citenamefont {Mun},
  \citenamefont {Caudle}, \citenamefont {Harmon}, \citenamefont {Bud’ko},
  \citenamefont {Canfield},\ and\ \citenamefont {Kaminski}}]{Liu2011}%
  \BibitemOpen
  \bibfield  {author} {\bibinfo {author} {\bibfnamefont {C.}~\bibnamefont
  {Liu}}, \bibinfo {author} {\bibfnamefont {Y.}~\bibnamefont {Lee}}, \bibinfo
  {author} {\bibfnamefont {T.}~\bibnamefont {Kondo}}, \bibinfo {author}
  {\bibfnamefont {E.~D.}\ \bibnamefont {Mun}}, \bibinfo {author} {\bibfnamefont
  {M.}~\bibnamefont {Caudle}}, \bibinfo {author} {\bibfnamefont {B.~N.}\
  \bibnamefont {Harmon}}, \bibinfo {author} {\bibfnamefont {S.~L.}\
  \bibnamefont {Bud’ko}}, \bibinfo {author} {\bibfnamefont {P.~C.}\
  \bibnamefont {Canfield}}, \ and\ \bibinfo {author} {\bibfnamefont
  {A.}~\bibnamefont {Kaminski}},\ }\href {\doibase 10.1103/PhysRevB.83.205133}
  {\bibfield  {journal} {\bibinfo  {journal} {Phys. Rev. B}\ }\textbf {\bibinfo
  {volume} {83}},\ \bibinfo {pages} {205133} (\bibinfo {year}
  {2011})}\BibitemShut {NoStop}%
\bibitem [{\citenamefont {Kim}\ \emph {et~al.}(2016)\citenamefont {Kim},
  \citenamefont {Wang}, \citenamefont {Nakajima}, \citenamefont {Hu},
  \citenamefont {Ziemak}, \citenamefont {Syers}, \citenamefont {Wang},
  \citenamefont {Hodovanets}, \citenamefont {Denlinger}, \citenamefont
  {Brydon}, \citenamefont {Agterberg}, \citenamefont {Tanatar}, \citenamefont
  {Prozorov},\ and\ \citenamefont {Paglione}}]{Kim2016}%
  \BibitemOpen
  \bibfield  {author} {\bibinfo {author} {\bibfnamefont {H.}~\bibnamefont
  {Kim}}, \bibinfo {author} {\bibfnamefont {K.}~\bibnamefont {Wang}}, \bibinfo
  {author} {\bibfnamefont {Y.}~\bibnamefont {Nakajima}}, \bibinfo {author}
  {\bibfnamefont {R.}~\bibnamefont {Hu}}, \bibinfo {author} {\bibfnamefont
  {S.}~\bibnamefont {Ziemak}}, \bibinfo {author} {\bibfnamefont
  {P.}~\bibnamefont {Syers}}, \bibinfo {author} {\bibfnamefont
  {L.}~\bibnamefont {Wang}}, \bibinfo {author} {\bibfnamefont {H.}~\bibnamefont
  {Hodovanets}}, \bibinfo {author} {\bibfnamefont {J.~D.}\ \bibnamefont
  {Denlinger}}, \bibinfo {author} {\bibfnamefont {P.~M.~R.}\ \bibnamefont
  {Brydon}}, \bibinfo {author} {\bibfnamefont {D.~F.}\ \bibnamefont
  {Agterberg}}, \bibinfo {author} {\bibfnamefont {M.~A.}\ \bibnamefont
  {Tanatar}}, \bibinfo {author} {\bibfnamefont {R.}~\bibnamefont {Prozorov}}, \
  and\ \bibinfo {author} {\bibfnamefont {J.}~\bibnamefont {Paglione}},\
  }\href@noop {} {\enquote {\bibinfo {title} {Beyond spin-triplet: Nodal
  topological superconductivity in a noncentrosymmetric semimetal},}\ }
  (\bibinfo {year} {2016}),\ \Eprint {http://arxiv.org/abs/1603.03375}
  {arXiv:1603.03375 [cond-mat.supr-con]} \BibitemShut {NoStop}%
\bibitem [{\citenamefont {Nakajima}\ \emph {et~al.}(2016)\citenamefont
  {Nakajima}, \citenamefont {Hu}, \citenamefont {Kirshenbaum}, \citenamefont
  {Hughes}, \citenamefont {Syers}, \citenamefont {Wang}, \citenamefont {Wang},
  \citenamefont {Wang}, \citenamefont {Saha}, \citenamefont {Pratt},
  \citenamefont {Lynn},\ and\ \citenamefont {Paglione}}]{Nakajima2016}%
  \BibitemOpen
  \bibfield  {author} {\bibinfo {author} {\bibfnamefont {Y.}~\bibnamefont
  {Nakajima}}, \bibinfo {author} {\bibfnamefont {R.}~\bibnamefont {Hu}},
  \bibinfo {author} {\bibfnamefont {K.}~\bibnamefont {Kirshenbaum}}, \bibinfo
  {author} {\bibfnamefont {A.}~\bibnamefont {Hughes}}, \bibinfo {author}
  {\bibfnamefont {P.}~\bibnamefont {Syers}}, \bibinfo {author} {\bibfnamefont
  {X.}~\bibnamefont {Wang}}, \bibinfo {author} {\bibfnamefont {K.}~\bibnamefont
  {Wang}}, \bibinfo {author} {\bibfnamefont {R.}~\bibnamefont {Wang}}, \bibinfo
  {author} {\bibfnamefont {S.~R.}\ \bibnamefont {Saha}}, \bibinfo {author}
  {\bibfnamefont {D.}~\bibnamefont {Pratt}}, \bibinfo {author} {\bibfnamefont
  {J.~W.}\ \bibnamefont {Lynn}}, \ and\ \bibinfo {author} {\bibfnamefont
  {J.}~\bibnamefont {Paglione}},\ }\href {\doibase 10.1126/sciadv.1500242}
  {\bibfield  {journal} {\bibinfo  {journal} {Sci. Adv.}\ }\textbf {\bibinfo
  {volume} {1}},\ \bibinfo {pages} {e1500242} (\bibinfo {year}
  {2016})}\BibitemShut {NoStop}%
\bibitem [{\citenamefont {Ali}\ \emph {et~al.}(2014)\citenamefont {Ali},
  \citenamefont {Gibson}, \citenamefont {Klimczuk},\ and\ \citenamefont
  {Cava}}]{Ali2014}%
  \BibitemOpen
  \bibfield  {author} {\bibinfo {author} {\bibfnamefont {M.~N.}\ \bibnamefont
  {Ali}}, \bibinfo {author} {\bibfnamefont {Q.~D.}\ \bibnamefont {Gibson}},
  \bibinfo {author} {\bibfnamefont {T.}~\bibnamefont {Klimczuk}}, \ and\
  \bibinfo {author} {\bibfnamefont {R.~J.}\ \bibnamefont {Cava}},\ }\href
  {\doibase 10.1103/PhysRevB.89.020505} {\bibfield  {journal} {\bibinfo
  {journal} {Phys. Rev. B}\ }\textbf {\bibinfo {volume} {89}},\ \bibinfo
  {pages} {020505} (\bibinfo {year} {2014})}\BibitemShut {NoStop}%
\bibitem [{\citenamefont {Bian}\ \emph {et~al.}(2016)\citenamefont {Bian},
  \citenamefont {Chang}, \citenamefont {Sankar}, \citenamefont {Xu},
  \citenamefont {Zheng}, \citenamefont {Neupert}, \citenamefont {Chiu},
  \citenamefont {Huang}, \citenamefont {Chang}, \citenamefont {Belopolski},
  \citenamefont {Sanchez}, \citenamefont {Neupane}, \citenamefont {Alidoust},
  \citenamefont {Liu}, \citenamefont {Wang}, \citenamefont {Lee}, \citenamefont
  {Jeng}, \citenamefont {Zhang}, \citenamefont {Yuan}, \citenamefont {Jia},
  \citenamefont {Bansil}, \citenamefont {Chou}, \citenamefont {Lin},\ and\
  \citenamefont {Hasan}}]{Bian2016}%
  \BibitemOpen
  \bibfield  {author} {\bibinfo {author} {\bibfnamefont {G.}~\bibnamefont
  {Bian}}, \bibinfo {author} {\bibfnamefont {T.-R.}\ \bibnamefont {Chang}},
  \bibinfo {author} {\bibfnamefont {R.}~\bibnamefont {Sankar}}, \bibinfo
  {author} {\bibfnamefont {S.-Y.}\ \bibnamefont {Xu}}, \bibinfo {author}
  {\bibfnamefont {H.}~\bibnamefont {Zheng}}, \bibinfo {author} {\bibfnamefont
  {T.}~\bibnamefont {Neupert}}, \bibinfo {author} {\bibfnamefont {C.-K.}\
  \bibnamefont {Chiu}}, \bibinfo {author} {\bibfnamefont {S.-M.}\ \bibnamefont
  {Huang}}, \bibinfo {author} {\bibfnamefont {G.}~\bibnamefont {Chang}},
  \bibinfo {author} {\bibfnamefont {I.}~\bibnamefont {Belopolski}}, \bibinfo
  {author} {\bibfnamefont {D.~S.}\ \bibnamefont {Sanchez}}, \bibinfo {author}
  {\bibfnamefont {M.}~\bibnamefont {Neupane}}, \bibinfo {author} {\bibfnamefont
  {N.}~\bibnamefont {Alidoust}}, \bibinfo {author} {\bibfnamefont
  {C.}~\bibnamefont {Liu}}, \bibinfo {author} {\bibfnamefont {B.}~\bibnamefont
  {Wang}}, \bibinfo {author} {\bibfnamefont {C.-C.}\ \bibnamefont {Lee}},
  \bibinfo {author} {\bibfnamefont {H.-T.}\ \bibnamefont {Jeng}}, \bibinfo
  {author} {\bibfnamefont {C.}~\bibnamefont {Zhang}}, \bibinfo {author}
  {\bibfnamefont {Z.}~\bibnamefont {Yuan}}, \bibinfo {author} {\bibfnamefont
  {S.}~\bibnamefont {Jia}}, \bibinfo {author} {\bibfnamefont {A.}~\bibnamefont
  {Bansil}}, \bibinfo {author} {\bibfnamefont {F.}~\bibnamefont {Chou}},
  \bibinfo {author} {\bibfnamefont {H.}~\bibnamefont {Lin}}, \ and\ \bibinfo
  {author} {\bibfnamefont {M.~Z.}\ \bibnamefont {Hasan}},\ }\href {\doibase
  10.1038/ncomms10556} {\bibfield  {journal} {\bibinfo  {journal} {Nat.
  Commun.}\ }\textbf {\bibinfo {volume} {7}},\ \bibinfo {pages} {10556}
  (\bibinfo {year} {2016})}\BibitemShut {NoStop}%
\bibitem [{\citenamefont {Guan}\ \emph {et~al.}(2016)\citenamefont {Guan},
  \citenamefont {Chen}, \citenamefont {Chu}, \citenamefont {Sankar},
  \citenamefont {Chou}, \citenamefont {Jeng}, \citenamefont {Chang},\ and\
  \citenamefont {Chuang}}]{Guan2016}%
  \BibitemOpen
  \bibfield  {author} {\bibinfo {author} {\bibfnamefont {S.-Y.}\ \bibnamefont
  {Guan}}, \bibinfo {author} {\bibfnamefont {P.-J.}\ \bibnamefont {Chen}},
  \bibinfo {author} {\bibfnamefont {M.-W.}\ \bibnamefont {Chu}}, \bibinfo
  {author} {\bibfnamefont {R.}~\bibnamefont {Sankar}}, \bibinfo {author}
  {\bibfnamefont {F.}~\bibnamefont {Chou}}, \bibinfo {author} {\bibfnamefont
  {H.-T.}\ \bibnamefont {Jeng}}, \bibinfo {author} {\bibfnamefont {C.-S.}\
  \bibnamefont {Chang}}, \ and\ \bibinfo {author} {\bibfnamefont {T.-M.}\
  \bibnamefont {Chuang}},\ }\href@noop {} {\enquote {\bibinfo {title}
  {Superconducting topological surface states in non-centrosymmetric bulk
  superconductor {PbTaSe$_2$}},}\ } (\bibinfo {year} {2016}),\ \Eprint
  {http://arxiv.org/abs/1605.00548} {arXiv:1605.00548 [cond-mat.supr-con]}
  \BibitemShut {NoStop}%
\bibitem [{\citenamefont {Neupane}\ \emph {et~al.}(2015)\citenamefont
  {Neupane}, \citenamefont {Alidoust}, \citenamefont {Xu}, \citenamefont
  {Belopolski}, \citenamefont {Sanchez}, \citenamefont {Chang}, \citenamefont
  {Jeng}, \citenamefont {Lin}, \citenamefont {Bansil}, \citenamefont
  {Kaczorowski}, \citenamefont {Hasan},\ and\ \citenamefont
  {Durakiewicz}}]{Neupane1505}%
  \BibitemOpen
  \bibfield  {author} {\bibinfo {author} {\bibfnamefont {M.}~\bibnamefont
  {Neupane}}, \bibinfo {author} {\bibfnamefont {N.}~\bibnamefont {Alidoust}},
  \bibinfo {author} {\bibfnamefont {S.-Y.}\ \bibnamefont {Xu}}, \bibinfo
  {author} {\bibfnamefont {I.}~\bibnamefont {Belopolski}}, \bibinfo {author}
  {\bibfnamefont {D.~S.}\ \bibnamefont {Sanchez}}, \bibinfo {author}
  {\bibfnamefont {T.-R.}\ \bibnamefont {Chang}}, \bibinfo {author}
  {\bibfnamefont {H.-T.}\ \bibnamefont {Jeng}}, \bibinfo {author}
  {\bibfnamefont {H.}~\bibnamefont {Lin}}, \bibinfo {author} {\bibfnamefont
  {A.}~\bibnamefont {Bansil}}, \bibinfo {author} {\bibfnamefont
  {D.}~\bibnamefont {Kaczorowski}}, \bibinfo {author} {\bibfnamefont {M.~Z.}\
  \bibnamefont {Hasan}}, \ and\ \bibinfo {author} {\bibfnamefont
  {T.}~\bibnamefont {Durakiewicz}},\ }\href@noop {} {\enquote {\bibinfo {title}
  {Discovery of the topological surface state in a noncentrosymmetric
  superconductor {BiPd}},}\ } (\bibinfo {year} {2015}),\ \Eprint
  {http://arxiv.org/abs/1505.03466} {arXiv:1505.03466 [cond-mat.mes-hall]}
  \BibitemShut {NoStop}%
\bibitem [{\citenamefont {Peets}(2014)}]{PeetsLT27}%
  \BibitemOpen
  \bibfield  {author} {\bibinfo {author} {\bibfnamefont {D.~C.}\ \bibnamefont
  {Peets}},\ }\href {\doibase 10.1088/1742-6596/568/2/022037} {\bibfield
  {journal} {\bibinfo  {journal} {J. Phys.: Conf. Ser.}\ }\textbf {\bibinfo
  {volume} {568}},\ \bibinfo {pages} {022037} (\bibinfo {year}
  {2014})}\BibitemShut {NoStop}%
\bibitem [{sup()}]{supp}%
  \BibitemOpen
  \href@noop {} {}\bibinfo {note} {See Supplemental Material at
  http://link.aps.org/ supplemental/10.1103/$\ldots$ for additional ARPES
  measurements.}\BibitemShut {Stop}%
\bibitem [{\citenamefont {Singh}\ \emph {et~al.}(2013)\citenamefont {Singh},
  \citenamefont {Enayat}, \citenamefont {White},\ and\ \citenamefont
  {Wahl}}]{Singh2013}%
  \BibitemOpen
  \bibfield  {author} {\bibinfo {author} {\bibfnamefont {U.}~\bibnamefont
  {Singh}}, \bibinfo {author} {\bibfnamefont {M.}~\bibnamefont {Enayat}},
  \bibinfo {author} {\bibfnamefont {S.}~\bibnamefont {White}}, \ and\ \bibinfo
  {author} {\bibfnamefont {P.}~\bibnamefont {Wahl}},\ }\href {\doibase
  10.1063/1.4788941} {\bibfield  {journal} {\bibinfo  {journal} {Rev. Sci.
  Instr.}\ }\textbf {\bibinfo {volume} {84}},\ \bibinfo {pages} {013708}
  (\bibinfo {year} {2013})}\BibitemShut {NoStop}%
\bibitem [{\citenamefont {Andersen}(1975)}]{andersenPRB75}%
  \BibitemOpen
  \bibfield  {author} {\bibinfo {author} {\bibfnamefont {O.~K.}\ \bibnamefont
  {Andersen}},\ }\href {\doibase 10.1103/PhysRevB.12.3060} {\bibfield
  {journal} {\bibinfo  {journal} {Phys. Rev. B}\ }\textbf {\bibinfo {volume}
  {12}},\ \bibinfo {pages} {3060} (\bibinfo {year} {1975})}\BibitemShut
  {NoStop}%
\bibitem [{\citenamefont {Antonov}\ \emph {et~al.}(2004)\citenamefont
  {Antonov}, \citenamefont {Harmon},\ and\ \citenamefont
  {Yaresko}}]{book:AHY04}%
  \BibitemOpen
  \bibfield  {author} {\bibinfo {author} {\bibfnamefont {V.}~\bibnamefont
  {Antonov}}, \bibinfo {author} {\bibfnamefont {B.}~\bibnamefont {Harmon}}, \
  and\ \bibinfo {author} {\bibfnamefont {A.}~\bibnamefont {Yaresko}},\
  }\href@noop {} {\emph {\bibinfo {title} {Electronic structure and
  magneto-optical properties of solids}}}\ (\bibinfo  {publisher} {Kluwer
  Academic Publishers},\ \bibinfo {address} {Dordrecht, Boston, London},\
  \bibinfo {year} {2004})\BibitemShut {NoStop}%
\bibitem [{\citenamefont {Perlov}\ \emph {et~al.}()\citenamefont {Perlov},
  \citenamefont {Yaresko},\ and\ \citenamefont {Antonov}}]{perlovYaresko}%
  \BibitemOpen
  \bibfield  {author} {\bibinfo {author} {\bibfnamefont {A.~Y.}\ \bibnamefont
  {Perlov}}, \bibinfo {author} {\bibfnamefont {A.~N.}\ \bibnamefont {Yaresko}},
  \ and\ \bibinfo {author} {\bibfnamefont {V.~N.}\ \bibnamefont {Antonov}},\
  }\href@noop {} {}\bibinfo {note} {PY-LMTO: A Spin-Polarized Relativistic LMTO
  Package for Electronic Structure Calculations (unpublished)}\BibitemShut
  {NoStop}%
\bibitem [{\citenamefont {Macdonald}\ \emph {et~al.}(1980)\citenamefont
  {Macdonald}, \citenamefont {Pickett},\ and\ \citenamefont
  {Koelling}}]{MPK80}%
  \BibitemOpen
  \bibfield  {author} {\bibinfo {author} {\bibfnamefont {A.~H.}\ \bibnamefont
  {Macdonald}}, \bibinfo {author} {\bibfnamefont {W.~E.}\ \bibnamefont
  {Pickett}}, \ and\ \bibinfo {author} {\bibfnamefont {D.~D.}\ \bibnamefont
  {Koelling}},\ }\href {\doibase 10.1088/0022-3719/13/14/009} {\bibfield
  {journal} {\bibinfo  {journal} {J. Phys. C}\ }\textbf {\bibinfo {volume}
  {13}},\ \bibinfo {pages} {2675} (\bibinfo {year} {1980})}\BibitemShut
  {NoStop}%
\bibitem [{\citenamefont {Bychkov}\ and\ \citenamefont
  {Rashba}(1984)}]{Bychkov1984}%
  \BibitemOpen
  \bibfield  {author} {\bibinfo {author} {\bibfnamefont {Y.~A.}\ \bibnamefont
  {Bychkov}}\ and\ \bibinfo {author} {\bibfnamefont {{\'E}.~I.}\ \bibnamefont
  {Rashba}},\ }\href {http://www.jetpletters.ac.ru/ps/1264/article_19121.shtml}
  {\bibfield  {journal} {\bibinfo  {journal} {JETP Lett.}\ }\textbf {\bibinfo
  {volume} {39}},\ \bibinfo {pages} {78} (\bibinfo {year} {1984})}\BibitemShut
  {NoStop}%
\bibitem [{\citenamefont {LaShell}\ \emph {et~al.}(1996)\citenamefont
  {LaShell}, \citenamefont {McDougall},\ and\ \citenamefont
  {Jensen}}]{LaShell1996}%
  \BibitemOpen
  \bibfield  {author} {\bibinfo {author} {\bibfnamefont {S.}~\bibnamefont
  {LaShell}}, \bibinfo {author} {\bibfnamefont {B.~A.}\ \bibnamefont
  {McDougall}}, \ and\ \bibinfo {author} {\bibfnamefont {E.}~\bibnamefont
  {Jensen}},\ }\href {\doibase 10.1103/PhysRevLett.77.3419} {\bibfield
  {journal} {\bibinfo  {journal} {Phys. Rev. Lett.}\ }\textbf {\bibinfo
  {volume} {77}},\ \bibinfo {pages} {3419} (\bibinfo {year}
  {1996})}\BibitemShut {NoStop}%
\bibitem [{\citenamefont {Koroteev}\ \emph {et~al.}(2004)\citenamefont
  {Koroteev}, \citenamefont {Bihlmayer}, \citenamefont {Gayone}, \citenamefont
  {Chulkov}, \citenamefont {Bl{\"u}gel}, \citenamefont {Echenique},\ and\
  \citenamefont {Hofmann}}]{Koroteev2004}%
  \BibitemOpen
  \bibfield  {author} {\bibinfo {author} {\bibfnamefont {Y.~M.}\ \bibnamefont
  {Koroteev}}, \bibinfo {author} {\bibfnamefont {G.}~\bibnamefont {Bihlmayer}},
  \bibinfo {author} {\bibfnamefont {J.~E.}\ \bibnamefont {Gayone}}, \bibinfo
  {author} {\bibfnamefont {E.~V.}\ \bibnamefont {Chulkov}}, \bibinfo {author}
  {\bibfnamefont {S.}~\bibnamefont {Bl{\"u}gel}}, \bibinfo {author}
  {\bibfnamefont {P.~M.}\ \bibnamefont {Echenique}}, \ and\ \bibinfo {author}
  {\bibfnamefont {P.}~\bibnamefont {Hofmann}},\ }\href {\doibase
  10.1103/PhysRevLett.93.046403} {\bibfield  {journal} {\bibinfo  {journal}
  {Phys. Rev. Lett.}\ }\textbf {\bibinfo {volume} {93}},\ \bibinfo {pages}
  {046403} (\bibinfo {year} {2004})}\BibitemShut {NoStop}%
\bibitem [{\citenamefont {Santos-Cottin}\ \emph {et~al.}(2016)\citenamefont
  {Santos-Cottin}, \citenamefont {Casula}, \citenamefont {Lantz}, \citenamefont
  {Klein}, \citenamefont {Petaccia}, \citenamefont {{Le F{\`e}vre}},
  \citenamefont {Bertran}, \citenamefont {Papalazarou}, \citenamefont {Marsi},\
  and\ \citenamefont {Gauzzi}}]{Santos2016}%
  \BibitemOpen
  \bibfield  {author} {\bibinfo {author} {\bibfnamefont {D.}~\bibnamefont
  {Santos-Cottin}}, \bibinfo {author} {\bibfnamefont {M.}~\bibnamefont
  {Casula}}, \bibinfo {author} {\bibfnamefont {G.}~\bibnamefont {Lantz}},
  \bibinfo {author} {\bibfnamefont {Y.}~\bibnamefont {Klein}}, \bibinfo
  {author} {\bibfnamefont {L.}~\bibnamefont {Petaccia}}, \bibinfo {author}
  {\bibfnamefont {P.}~\bibnamefont {{Le F{\`e}vre}}}, \bibinfo {author}
  {\bibfnamefont {F.}~\bibnamefont {Bertran}}, \bibinfo {author} {\bibfnamefont
  {E.}~\bibnamefont {Papalazarou}}, \bibinfo {author} {\bibfnamefont
  {M.}~\bibnamefont {Marsi}}, \ and\ \bibinfo {author} {\bibfnamefont
  {A.}~\bibnamefont {Gauzzi}},\ }\href {\doibase 10.1038/ncomms11258}
  {\bibfield  {journal} {\bibinfo  {journal} {Nat. Commun.}\ }\textbf {\bibinfo
  {volume} {7}},\ \bibinfo {pages} {11258} (\bibinfo {year}
  {2016})}\BibitemShut {NoStop}%
\bibitem [{\citenamefont {Gierz}\ \emph {et~al.}(2010)\citenamefont {Gierz},
  \citenamefont {Stadtm\"{u}ller}, \citenamefont {Vuorinen}, \citenamefont
  {Lindroos}, \citenamefont {Meier}, \citenamefont {Dil}, \citenamefont
  {Kern},\ and\ \citenamefont {Ast}}]{gierz_structural_2010}%
  \BibitemOpen
  \bibfield  {author} {\bibinfo {author} {\bibfnamefont {I.}~\bibnamefont
  {Gierz}}, \bibinfo {author} {\bibfnamefont {B.}~\bibnamefont
  {Stadtm\"{u}ller}}, \bibinfo {author} {\bibfnamefont {J.}~\bibnamefont
  {Vuorinen}}, \bibinfo {author} {\bibfnamefont {M.}~\bibnamefont {Lindroos}},
  \bibinfo {author} {\bibfnamefont {F.}~\bibnamefont {Meier}}, \bibinfo
  {author} {\bibfnamefont {J.~H.}\ \bibnamefont {Dil}}, \bibinfo {author}
  {\bibfnamefont {K.}~\bibnamefont {Kern}}, \ and\ \bibinfo {author}
  {\bibfnamefont {C.~R.}\ \bibnamefont {Ast}},\ }\href {\doibase
  10.1103/PhysRevB.81.245430} {\bibfield  {journal} {\bibinfo  {journal}
  {Physical Review B}\ }\textbf {\bibinfo {volume} {81}},\ \bibinfo {pages}
  {245430} (\bibinfo {year} {2010})}\BibitemShut {NoStop}%
\bibitem [{\citenamefont {Bian}\ \emph {et~al.}(2013)\citenamefont {Bian},
  \citenamefont {Wang}, \citenamefont {Miller},\ and\ \citenamefont
  {Chiang}}]{bian_origin_2013}%
  \BibitemOpen
  \bibfield  {author} {\bibinfo {author} {\bibfnamefont {G.}~\bibnamefont
  {Bian}}, \bibinfo {author} {\bibfnamefont {X.}~\bibnamefont {Wang}}, \bibinfo
  {author} {\bibfnamefont {T.}~\bibnamefont {Miller}}, \ and\ \bibinfo {author}
  {\bibfnamefont {T.-C.}\ \bibnamefont {Chiang}},\ }\href {\doibase
  10.1103/PhysRevB.88.085427} {\bibfield  {journal} {\bibinfo  {journal}
  {Physical Review B}\ }\textbf {\bibinfo {volume} {88}},\ \bibinfo {pages}
  {085427} (\bibinfo {year} {2013})}\BibitemShut {NoStop}%
\bibitem [{\citenamefont {Yan}\ \emph {et~al.}(2016)\citenamefont {Yan},
  \citenamefont {Xu}, \citenamefont {He}, \citenamefont {Dong}, \citenamefont
  {Cho}, \citenamefont {Peets}, \citenamefont {Park},\ and\ \citenamefont
  {Li}}]{Yan2016}%
  \BibitemOpen
  \bibfield  {author} {\bibinfo {author} {\bibfnamefont {X.~B.}\ \bibnamefont
  {Yan}}, \bibinfo {author} {\bibfnamefont {Y.}~\bibnamefont {Xu}}, \bibinfo
  {author} {\bibfnamefont {L.~P.}\ \bibnamefont {He}}, \bibinfo {author}
  {\bibfnamefont {J.~K.}\ \bibnamefont {Dong}}, \bibinfo {author}
  {\bibfnamefont {H.~B.}\ \bibnamefont {Cho}}, \bibinfo {author} {\bibfnamefont
  {D.~C.}\ \bibnamefont {Peets}}, \bibinfo {author} {\bibfnamefont {J.-G.}\
  \bibnamefont {Park}}, \ and\ \bibinfo {author} {\bibfnamefont {S.~Y.}\
  \bibnamefont {Li}},\ }\href {\doibase 10.1088/0953-2048/29/6/065001}
  {\bibfield  {journal} {\bibinfo  {journal} {Supercond. Sci. Tech.}\ }\textbf
  {\bibinfo {volume} {29}},\ \bibinfo {pages} {065001} (\bibinfo {year}
  {2016})}\BibitemShut {NoStop}%
\end{thebibliography}%

\clearpage

\setcounter{figure}{0}
\renewcommand{\thefigure}{S\arabic{figure}}


\begin{center}
\textbf{
  \large{Supplementary  Information for\\
  ``Observation of Dirac surface states in the noncentrosymmetric superconductor BiPd"}
}
\vspace{0.1cm}

H.~M.~Benia, E.~Rampi, C.~Trainer, C.~M.~Yim, A.~Maldonado, D.~C.~Peets, A.~St\"ohr, U.~Starke, K.~Kern, A.~Yaresko, G.~Levy, A.~Damascelli, C.~R.~Ast, A.~P.~Schnyder, and P.~Wahl
\end{center}

\vspace{0.2cm}


\begin{figure*}[thb]
\includegraphics[width=\textwidth]{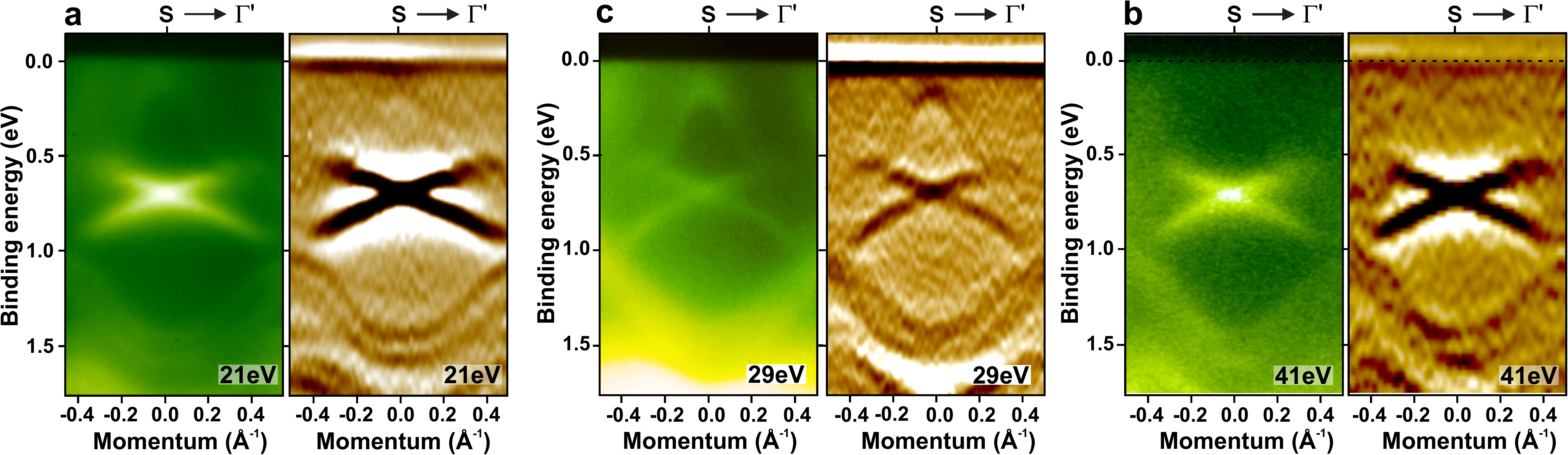}
\caption{\label{mSFig1}Experimental surface electronic band structure of BiPd(010) along the S--$\Gamma^{\prime}$ direction and second derivative images, measured using different photon energies: (a) 21~eV (He-I), (b) 29~eV (synchrotron light), and (c)  42~eV (He-II). Changing the incident photon energy changes $k_z$, but these bands do not disperse, as would be expected for two-dimensional surface states.}
\end{figure*}

\begin{figure*}[thb]
\includegraphics[width=\textwidth]{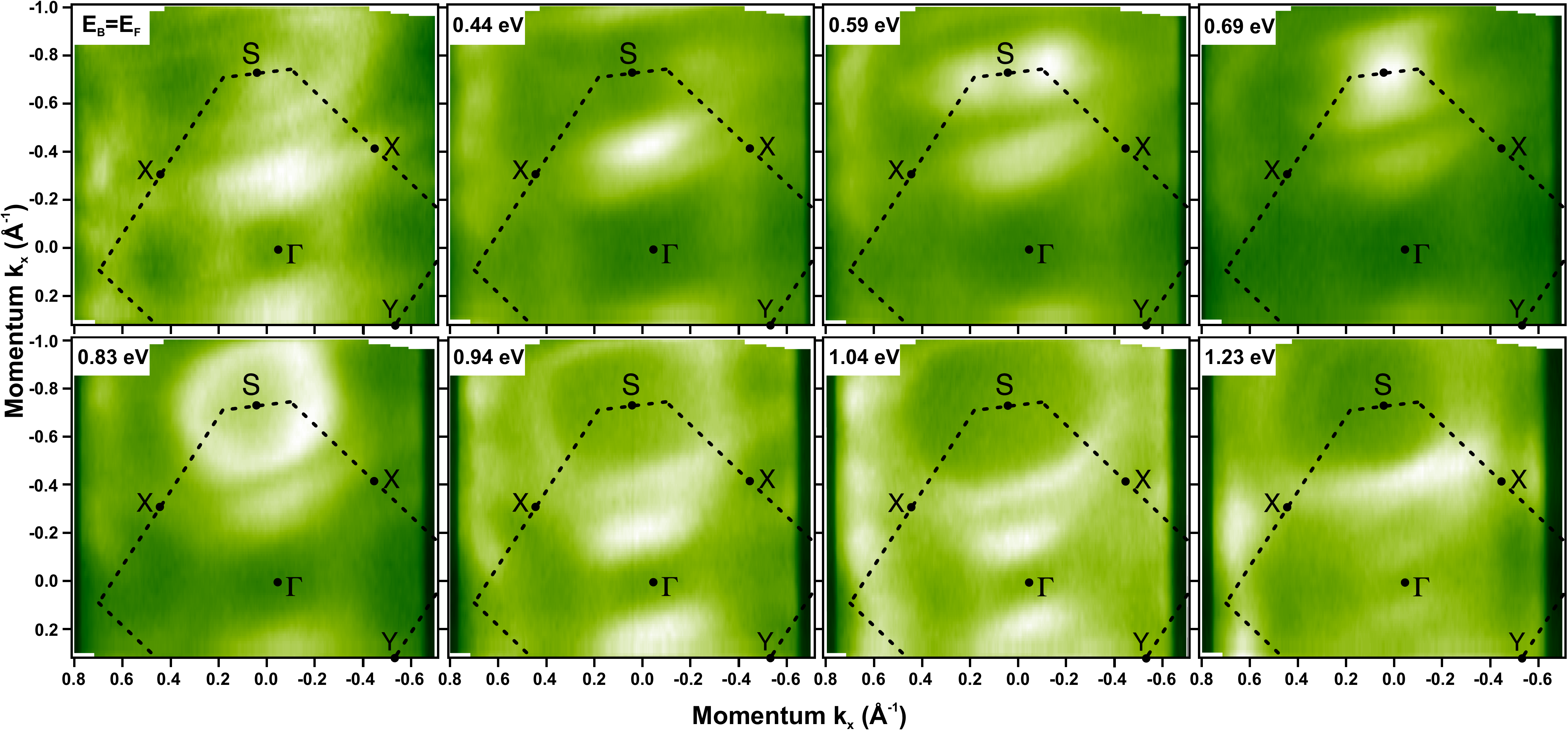}
\caption{\label{mSFig2}Additional constant energy cuts at different binding energies $E_B$ of the experimental surface electronic band structure of BiPd(010). ARPES measurements have been performed using a He-I light source (21~eV). The dashed line represents the projected surface Brillouin zone.}
\end{figure*}

Figures~\ref{mSFig1} and \ref{mSFig2} show supporting ARPES data. In Fig.~\ref{mSFig1} ARPES intensity maps are displayed for three different incident photon energies.  These energies probe different $k_z$, leading to rather different bulk band structures in this relatively three-dimensional material, but the two-dimensional surface states do not disperse. This figure also includes second-derivative images, showing the surface states with greater contrast.  Figure~\ref{mSFig2} shows additional constant-energy cuts for an incident photon energy of 21~eV, fleshing out the dispersion of the surface states in greater detail.

\begin{figure*}[thb]
\includegraphics[width=\textwidth]{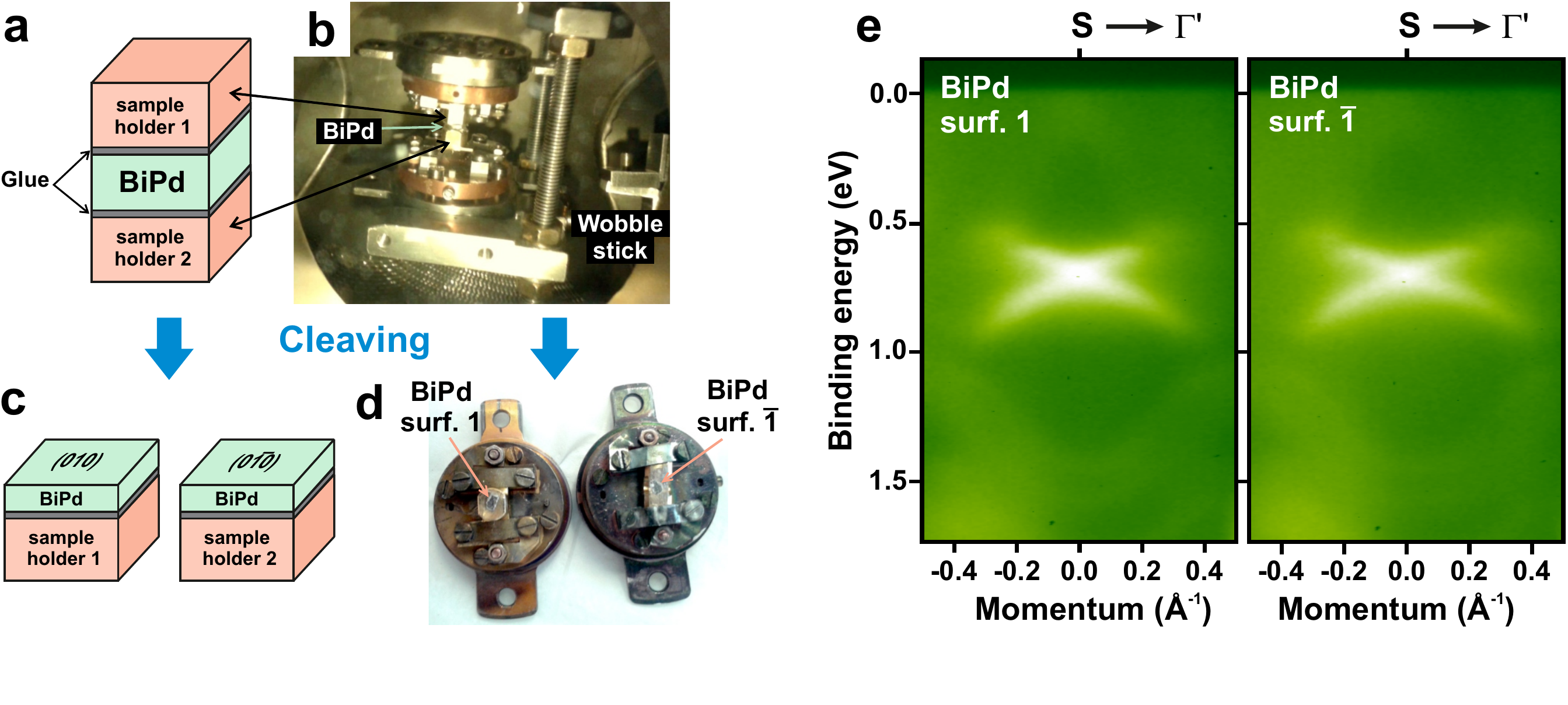}
\caption{\label{mSFig3}
{(a) Schematic drawing of the experimental design shown in the digital photograph in (b) of a BiPd crystal put into a sandwich arrangement between two sample holders. (c) Schematic drawing showing an ideal situation, where the noncentrosymmetric BiPd is a perfect single crystal without twinning. In this case cleaving results in two cleaved surfaces characterized by a single termination/orientation each. (d) Digital photograph showing the two sample holders with the resulting two BiPd crystals after cleaving in UHV and ARPES measurements. (e) Experimental band structures measured successively on the two cleaved  surfaces at 100 K using a He-I lab source. Since the ARPES data of the two cleaved surfaces exhibit almost identical features, we conclude that the crystals are composed of
twin domains, whose size is smaller then the ARPES beam spot.}
}
\end{figure*}

Figure~\ref{mSFig3} shows the cleaving setup that was used to obtain  surfaces with opposite surface termination. Since the ARPES measurements of the two cleaved surfaces exhibit similar features, we conclude that the BiPd crystal consists of twin domains,
with the twin domain size  smaller than the ARPES beam spot.

\end{document}